\documentclass[iop]{emulateapj}
\usepackage[T1]{fontenc}
\usepackage{amsmath}
\usepackage{amssymb}	
\usepackage{graphicx,aas_macros}
\usepackage{subfigure}
\usepackage[breaklinks,colorlinks,citecolor=blue]{hyperref}
\usepackage[all]{hypcap}
\usepackage{bm}
\usepackage{color}
\usepackage[dvipsnames]{xcolor}
\usepackage{setspace}

\def\stacksymbols #1#2#3#4{\def\theguybelow{#2}
        \def\verticalposition{\lower#3pt}
        \def\spacingwithinsymbol{\baselineskip0pt\lineskip#4pt}
        \mathrel{\mathpalette\intermediary#1}}
\def\intermediary #1#2{\verticalposition\vbox{\spacingwithinsymbol
        \everycr={}\tabskip0pt
        \halign{$\mathsurround0pt#1\hfil##\hfil$\crcr#2\crcr
                \theguybelow\crcr}}}

\def\lta{\stacksymbols{<}{\sim}{2.5}{.2}}
\def\gta{\stacksymbols{>}{\sim}{2.5}{.2}}

\def\kms{{\rm km\:s^{-1}}}

\newcommand{\be}{\begin{equation}}
\newcommand{\ee}{\end{equation}}
\newcommand{\bea}{\begin{eqnarray}}
\newcommand{\eea}{\end{eqnarray}}

\newcommand{\msun}{M_{\odot}}

\newcommand{\cc}{{\rm cm}^{-3}}

\begin{document}

\shortauthors{M.~Gaspari et al.}
\title{ 
Shaken Snow Globes:
Kinematic Tracers of the Multiphase Condensation Cascade in Massive Galaxies, Groups, and Clusters}
\shorttitle{Constraining hot gas turbulence from the warm and cold phase}
\author{M.~Gaspari$^{1,*,\dagger}$, M.~McDonald$^2$, S.~L.~Hamer$^3$, F.~Brighenti$^4$, P.~Temi$^5$, M.~Gendron-Marsolais$^6$, J.~Hlavacek-Larrondo$^6$, A.~C.~Edge$^7$, N.~Werner$^{8,9,10}$, P.~Tozzi$^{11}$, M.~Sun$^{12}$, J.~M.~Stone$^1$, G.~R.~Tremblay$^{13}$, M.~T.~Hogan$^{14}$, D.~Eckert$^{15}$, S.~Ettori$^{16,17}$, H.~Yu$^{18}$, V.~Biffi$^{19,20}$, S.~Planelles$^{21}$
\vspace{+0.1cm}
}
\affil{\scriptsize$^1\,$Department of Astrophysical Sciences, Princeton University, 4 Ivy Lane, Princeton, NJ 08544-1001, USA \\
 $^2\,$Kavli Institute for Astrophysics and Space Research, MIT, Cambridge, MA 02139, USA \\ 
 $^3\,$CRAL, Lyon Observatory, CNRS, Universit\'e Lyon 1, 9 Avenue Charles Andr\'e, F-69561 Saint Genis-Laval, France \\
 $^4\,$Astronomy Department, University of Bologna, Via Piero Gobetti, 93/3, 40129 Bologna, Italy\\
 $^5\,$Astrophysics Branch, NASA Ames Research Center, Moffett Field, CA 94035, USA \\
 $^6\,$Department of Physics, University of Montreal, Montr\'eal, QC H3C 3J7, Canada\\
 $^7\,$Department of Physics, Durham University, Durham, DHL 3LE, United Kingdom\\
 $^8\,$MTA-E\"otv\"os University Lend\"ulet Hot Universe Research Group, P\'azm\'any P\'eter s\'et\'any 1/A, Budapest, 1117, Hungary \\
$^9\,$Dep.~of Theoretical Physics and Astrophysics, Faculty of Science, Masaryk University, Kotl\'a\v{r}sk\'a 2, Brno, 611 37, Czech Republic \\
$^{10}\,$School of Science, Hiroshima University, 1-3-1 Kagamiyama, Higashi-Hiroshima 739-8526, Japan\\
$^{11}\,$INAF, Astronomy Observatory of Florence, Largo Enrico Fermi 5, 50125, Firenze, Italy\\
$^{12}\,$Physics Department, University of Alabama in Huntsville, Huntsville, AL 35899, USA\\
$^{13}\,$Harvard-Smithsonian Center for Astrophysics, 60 Garden Street, Cambridge, MA 02138, USA\\
$^{14}\,$Department of Physics and Astronomy, University of Waterloo, Waterloo, ON, N2L 3G1, Canada\\
$^{15}\,$Max Planck Institute for Extraterrestrial Physics, Giessenbachstr., 85741, Garching, Germany \\
$^{16}\,$INAF, Astronomy Observatory of Bologna, Via Piero Gobetti, 93/3, 40129 Bologna, Italy\\
$^{17}\,$INFN, Sezione di Bologna, viale Berti Pichat 6/2, 40127 Bologna, Italy\\
$^{18}\,$Department of Astronomy, Beijing Normal University, Beijing, 100875, China\\
$^{19}\,$Department of Physics and Astronomy, University of Trieste, via Tiepolo 11, 34131 Trieste, Italy\\
$^{20}\,$INAF, Astronomy Observatory of Trieste -- OATs, via Tiepolo 11, 34131 Trieste, Italy\\
$^{21}\,$Department of Astronomy and Astrophysics, University of Valencia, C/Dr. Moliner 50, 46100 Valencia, Spain
 }
\altaffiltext{\hspace{-0.15in} * }{E-mail: mgaspari@astro.princeton.edu}
\altaffiltext{\hspace{-0.15in} $\dagger$ }{{\it Einstein} and {\it Spitzer} Fellow}

\begin{abstract}
\noindent
We propose a novel method to constrain turbulence and bulk motions in massive galaxies, galaxy groups and clusters, exploring both simulations and observations. As emerged in the recent picture of the top-down multiphase condensation, the hot gaseous halos are tightly linked to all other phases in terms of cospatiality and thermodynamics. While hot halos ($\sim$\,$10^7$\;K) are perturbed by subsonic turbulence, warm ($\sim$\,$10^4$\;K) ionized  and neutral filaments condense out of the turbulent eddies. The peaks condense into cold molecular clouds ($< 100$\;K) raining in the core via chaotic cold accretion (CCA). We show all phases are tightly linked in terms of the {\it ensemble} (wide-aperture) velocity dispersion along the line of sight. The correlation arises in complementary long-term AGN feedback simulations and high-resolution CCA  runs, and is corroborated by the combined {\it Hitomi} and new Integral Field Unit measurements in Perseus cluster. The ensemble multiphase gas distributions (from UV to radio band) are characterized by substantial spectral line broadening ($\sigma_{v,{\rm los}}\approx100$\,-\,$200\;\kms$) with mild line shift. On the other hand, pencil-beam detections (as HI absorption against the AGN backlight) sample the small-scale clouds displaying smaller broadening and significant line shift up to several 100\;$\kms$ (for those falling toward the AGN), with increased scatter due to the turbulence intermittency. We present new ensemble $\sigma_{v,{\rm los}}$ of the warm H$\alpha$+[NII] gas in 72 observed cluster/group cores: the constraints are consistent with the simulations and can be used as robust proxies for the turbulent velocities, in particular for the challenging hot plasma (otherwise requiring extremely long X-ray exposures). Finally, we show the physically motivated criterion $C \equiv t_{\rm cool}/t_{\rm eddy}\approx 1$ best traces the condensation extent region and presence of multiphase gas in observed clusters and groups. The ensemble method can be applied to many available spectroscopic datasets and can substantially advance our understanding of multiphase halos in light of the next-generation multiwavelength missions.
\vspace{+0.22cm}
\end{abstract}


\section{Introduction} \label{s:intro}
\setcounter{footnote}{0}
\noindent
Despite our everyday solid-state experience, baryons populate the universe mostly in a diffuse gaseous form.
A new picture has recently emerged -- from both the theoretical and observational side -- describing the gaseous atmospheres of galaxies, groups, and clusters of galaxies.
While initially modeled as hydrostatic monophase systems, the gaseous halos filling the potential well of cosmic systems are complex atmospheres akin to Earth weather, following a top-down multiphase condensation cascade (e.g.,~\citealt{Gaspari:2017_cca,Gaspari:2017_uni}).
After falling at large redshift into the potential wells of dark matter halos, baryons heat up, forming hot plasma halos (intracluster, intragroup, and circumgalactic medium -- ICM, IGrM, CGM; \citealt{McNamara:2012,Sun:2012} for reviews). Such hot halos are the progenitors for other major condensed structures, including warm filaments, cold molecular clouds, and stellar/planetary systems.

During their evolution, the diffuse halos experience cyclical states, akin to the rapid alternation on Earth of sunny, cloudy, and rainy weather. From the thermal point of view, cosmic atmospheres span temperatures from several keV (${\rm 1\;keV} = 1.16\times10^7$\;K) of {\it hot} plasma halos to $T\sim10^4$\;K of {\it warm} ionized and neutral filaments to tens K of {\it cold} molecular clouds (as beautifully detected by ALMA), with particle number density on average anticorrelated with temperature ($n\sim10^{-3}$\,-\,$10^3\;\cc$). 
At the same time, from the dynamical point of view, cosmic atmospheres experience a continuous competition between chaotic turbulent motions and coherent rotational flows (turbulent Taylor number ${\rm Ta}<1$ or $>1$, respectively). 
Hotter, thermal pressure-supported halos often reside in the former chaotic regime due to the multiple drivers acting through cosmic time in a partially uncorrelated way: at larger radial distances (Mpc-scale) mergers and galaxy motions 
drive subsonic turbulence in the volume-filling phase
(e.g., \citealt{Vazza:2011,Miniati:2014,Khatri:2016}), while in the core ($r\lta50$\;kpc -- where the entropy profile slope changes) active galactic nucleus (AGN) feedback recurrently pumps energy via massive outflows and jets (e.g., \citealt{Lau:2017,Hillel:2017}); at the smallest scales, supernovae and stellar winds further preserve a minimum level of (compressive) turbulence (e.g., \citealt{Kim:2013}).

In the turbulent gaseous halos of clusters, groups, and galaxies (particularly massive ones), extended filaments and clouds condense out of the hot plasma in a {\it top-down} nonlinear\footnote{This nonlinear condensation process has significantly different properties from those of classic linear thermal instability (TI); the latter is mainly concerned with small overdensities overcoming buoyancy oscillations (e.g., \citealt{Field:1965,Balbus:1989,Burkert:2000,Pizzolato:2005,McCourt:2012} -- more in \S\ref{s:crit}).
} 
condensation cascade, forming a chaotic multiphase rain. 
The thermal state and kinematics of the progenitor hot plasma halo drive the formation and evolution of all the condensed structures, which inherit some of the parent properties.
Part of the inner condensed gas eventually accretes onto the central supermassive black hole (SMBH), igniting the feedback response and efficiently self-regulating the whole atmosphere over several Gyr (e.g., \citealt{Gaspari:2011a,Gaspari:2011b,Gaspari:2012b,Gaspari:2012a,Li:2014,Barai:2016,Yang:2016,Soker:2016,Meece:2017,Voit:2017}). 
This feeding process is known as {\it chaotic cold accretion} (CCA; \citealt{Gaspari:2013_cca})
and can intermittently boost the accretion rates up to $100\times$ the hot (Bondi) rate.
If turbulence is subdominant, the halo tends instead to condense in a disk structure (due to the preservation of angular momentum), reducing feeding and feedback -- this regime is more important for low-mass, spiral galaxies\footnote{The top-down rain differs from the bottom-up condensation in the disk of spiral galaxies, where the hot/warm phase is created in situ by supernovae which drive compressive, non-solenoidal turbulence (e.g., \citealt{McKee:1977,Kim:2013}). Nevertheless, the two complement each other, producing multiphase gas in the more extended halo and in the disk, respectively. Massive galaxies, groups, and clusters, lacking an extended disk (e.g., \citealt{Werner:2014}), typically reside in the top-down condensation regime.}.  
Finally, if the entropy of the halo (or cooling time) becomes too high, the whole atmosphere may simply prevent condensation and remain hot for an extended period of time, dramatically stifling the feedback response.
Overall, assessing the dynamical state of the multiphase halos is crucial to understand the past and predict the future evolution of cosmic structures. 

Although the thermal properties of gaseous halos are fairly well-constrained thanks to the last-generation X-ray, optical/IR, and radio telescopes (e.g., {\it Chandra}, XMM, {\it Hubble}, {\it Herschel}, and IRAM; \citealt{Combes:2007,McDonald:2010,McDonald:2011a,McNamara:2012,Canning:2013,Werner:2014,Tremblay:2015,Hamer:2016,Russell:2016,David:2017}), constraining the kinematics of the hot phase has proven to be very challenging, mainly due to the limited spectral resolution at high energies. 
The kinematics of the gas can be directly retrieved from the spectral line width (which is tied to the turbulent velocity dispersion) or the line centroid offset (which traces bulk motions). 
Recently, {\it Hitomi} gave us a sneak peek into the complexity of hot halos, finding $\simeq160\;\kms$ line-of-sight (LOS) velocity dispersions in the Perseus cluster core (\citealt{Hitomi:2016}). Turbulent motions can also be roughly estimated via relative plasma density fluctuations, which are related to the turbulent Mach number $\delta \rho/\rho\approx {\rm Ma_{1d}}$ (e.g., \citealt{Gaspari:2013_coma,Hofmann:2016,Eckert:2017_PS,Zhuravleva:2017}), finding subsonic Mach numbers in the ICM, although substructures contamination can introduce a significant noise. The subsonic turbulence is corroborated by the linewidth upper limits in combination with resonant scattering set by XMM-RGS (\citealt{Werner:2009,dePlaa:2012,Sanders:2013,Pinto:2015,Ogorzalek:2017}). Such a level of turbulence is also required to substantially suppress the emission measure in the soft X-ray spectrum (\citealt{Gaspari:2015_xspec}).

This paper continues our systematic investigation of the multiphase condensation and CCA mechanism (e.g., \citealt{Gaspari:2012a,Gaspari:2013_cca,Gaspari:2015_cca,Gaspari:2017_cca}), focusing on the gas kinematics.
By using state-of-the-art high-resolution hydrodynamic 3D simulations, complemented by new observations, we present a novel method to constrain the gas motions, taking advantage of the {\it ensemble}\footnote{Theoretically, meaning the global large-volume statistics of all of the condensed elements for a given phase; practically, referring to the use of (spectroscopic) observations with wide projected radial aperture $R$ ($\sim$\,arcmin or exceeding several kpc).} 
kinematics of the condensed multiphase filaments and clouds -- one of the most robust properties of the low-energy phases.
We will show that, singularly, each structure can take a large range of values of the random and bulk velocity components. Globally, and with enough statistics, the condensed structures can be considered as quasi-linear tracers of the turbulent eddies and cascade -- reminiscent of shaken snow globes. Vice versa, we can apply the same method to infer the kinematics of the cooler phase from the warm phase, or any different multiwavelength combination. As shown by the new observational constraints in \S\ref{s:obs}, this can be easily and efficiently leveraged by the Integral Field Unit (IFU) spectroscopy, which is advancing at a remarkable pace (e.g., MUSE, VIMOS, SITELLE).
In the other direction, small-aperture/`pencil-beam' (e.g., $R <$ a few 100 pc or $\sim$\,arcsec) detections -- such as HI absorption against the AGN backlight or CO emission -- can shed light on the mode of accretion onto SMBHs (e.g., CCA versus hot mode accretion; \citealt{David:2014,Hogan:2014,Tremblay:2016}) and on the properties of the small-scale clouds (e.g., \citealt{Maccagni:2017}).
As we live in an era of new exciting telescopes covering the radio and IR spectrum (e.g., ALMA, JWST, SKA, CARMA2) the proposed multiwavelength kinematics methods can be tested and used to advance our understanding of cosmic halos in galaxies, groups, and clusters.

Retrieving the velocity dispersion $\sigma_v$ of the hot halo opens up a simple and direct way to assess the presence of multiphase gas or ensuing condensation. A major debate in the recent literature concerns which is the best (and minimal) criterion to assess the condensation state of the hot halo as a function of characteristic timescales (e.g., \citealt{McCourt:2012,Sharma:2012,Gaspari:2012a,Voit:2015_nat,McNamara:2016,Hogan:2017}), including $t_{\rm cool}<1$\;Gyr, $t_{\rm cool}/t_{\rm ff}\lta 10$\,-\,30, or $t_{\rm cool}/t_{\rm cond}\lta 1$, where $t_{\rm cool}$, $t_{\rm ff}$, and $t_{\rm cond}$ are the cooling, gravitational, and conduction timescales, respectively. We will show that the crossing locus of the cooling time and the turbulent eddy time (which is a function of predominantly the ensemble gas velocity dispersion, $t_{\rm eddy} \propto \sigma_v^{-1}$, and directly accessible from observations) provides a robust criterion for the physical state of the hot halo, separating multiphase and monophase systems. 
Although the method can be applied to a large number of available datasets, it can also augment the next-generation X-ray missions (e.g., {\it Athena}, XARM, and {\it Lynx}) by providing precise and testable observables.

This work is structured as follows.
In \S\ref{s:sims}, we review the high-resolution 3D hydrodynamic simulations used in this study. 
In \S\ref{s:res}, we dissect the resulting correlations between all of the different phases (soft X-ray to UV/optical band to radio/21\,cm), in both long-term AGN feedback simulations (\S\ref{s:G12}) and CCA feeding simulations with pc-scale resolution (\S\ref{s:G17}). The main $\sigma_{v,\rm{los}}$ correlation is tested with the {\it Hitomi} and new SITELLE IFU direct measurements in the Perseus cluster. In \S\ref{s:lim}, we discuss the current limitations of the models and future improvements.
In \S\ref{s:obs}, we present new observational constraints -- together with available literature data -- obtained via the proposed ensemble (\S\ref{s:obs_ens}) and pencil-beam (\S\ref{s:obs_pen}) methods for the warm and cold gas in massive galaxies (many of which are central brightest cluster galaxies -- BCGs) and we compare them with the above numerical predictions.
In \S\ref{s:crit}, we discuss a key application of the ensemble measurement, 
presenting a new condensation criterion tied to the turbulence eddy turnover time for the presence and radial extension of the multiphase gas structures in clusters and groups. 
In \S\ref{s:conc}, we summarize the main results of the study and provide concluding remarks.\\

\section{Simulations} \label{s:sims}
\noindent
The core of the theoretical study (\S\ref{s:res}) is based on 3D hydrodynamic simulations (carried out with the Eulerian adaptive-mesh-refinement -- AMR -- code \texttt{FLASH4}),
combining them with new and recent multiwavelength observations (\S\ref{s:obs}).
We use as reference two complementary simulations, one covering the large-scale and long-term evolution, and the other covering the high-resolution and full multiphase cascade from the hot plasma to the molecular phase.
As we privileged high accuracy in space and time, the total computational cost is substantial, over 4 million CPU-hours. 
Since the simulations are unchanged compared with our previous investigations, 
we refer the interested reader to \citeauthor{Gaspari:2012a} (\citeyear{Gaspari:2012a} -- G12 hereafter) and \citeauthor{Gaspari:2017_cca} (\citeyear{Gaspari:2013_cca, Gaspari:2015_cca, Gaspari:2017_cca} -- G17 hereafter) for the details and nuances related to the modules and numerics adopted. Here we review the key features and relevant physics.\\

\subsection{G12 simulation: self-regulated AGN feedback} \label{s:G12m}

The goal of the G12 suite of simulations is to study the evolution and properties of the long-term self-regulated kinetic AGN feedback affecting the X-ray plasma halo. The simulation models a typical cool-core galaxy cluster with central plasma\footnote{The plasma average particle weight is $\mu\simeq 0.62$, with adiabatic index $\gamma =5/3$. The metal abundances are $Z\simeq0.3$\,-\,$1.0\;Z_\odot$ for the cluster and central galaxy, respectively.
} entropy $K_0\simeq15$\;keV\,cm$^2$ (minimum $t_{\rm cool}/t_{\rm ff} < 10$).
The plasma halo is initially perturbed by random fluctuations in density and temperature with 0.3 relative amplitude to model the presence of cosmic weather (G12; Sec.~2.3).
The maximum AMR resolution reaches 300\;pc, so that it is possible to evolve the system for several Gyr in a large $1.3^3$\;Mpc$^3$ domain. The static gravitational potential is dominated by the 
dark matter component with Navarro-Frenk-White profile; the cluster virial mass is $M_{\rm vir}\approx10^{15}\;\msun$ with gas fraction $\approx0.15$.

Besides hydrodynamics\footnote{In all runs, we employ the third-order accurate piecewise parabolic method (PPM) to solve the Euler hydrodynamics equations. Boundary conditions have all outflow permitted and inflow prohibited.}, 
the two key competing physics are radiative cooling and AGN feedback. 
The plasma radiative cooling
induces the gas to lose thermal energy, and thus pressure support, forming extended warm filaments via nonlinear condensation. 
The plasma halo emits radiation mainly via Bremsstrahlung above 1 keV and line recombination below such soft X-ray regime.
The emissivity is $\simeq n^2\,\Lambda(T,Z)$, where $\Lambda$ is the \citet{Sutherland:1993} plasma cooling function in collisional ionization equilibrium.
The plasma cooling curve incorporates calculations for H, He, C, N, O, Fe, Ne, Na, Si, Mg, Al, Ar, S, Cl, Ca, and Ni, and all stages of ionization.
Due to the limited resolution in this run,
condensation is halted at the warm phase regime around $10^4$\;K. The cooling source term is integrated with an exact solver with conservative time-step limiter (G12, Sec.~2.1).

Radiative cooling is counterbalanced by AGN feedback, in the form of massive subrelativistic outflows.
The bipolar outflow mass, momentum, and energy are injected through an internal boundary nozzle in the innermost resolved region (G12, Sec.~2.2) -- locus of the SMBH. The injected velocity is $v_{\rm out}=5\times10^4\;\kms$, which is typical of observed entrained FRI jets or ionized ultrafast outflows (e.g., \citealt{Tombesi:2013}). 
The injected kinetic power is self-regulated by the central inflow rate, $P_{\rm out}=(1/2)\dot M_{\rm out}\,v^2_{\rm out}\simeq \varepsilon \dot M_{\rm in}\,c^2$, where $\varepsilon=6\times10^{-3}$ is the mechanical efficiency able to avoid both overcooling and overheating. 
The triggering nuclear mass inflow $\dot M_{\rm in}$ ($r < 500$\;pc) is dominated by the condensed gas phase (thus linked to the cooling rate of the hot halo) in the form of raining filaments and clouds (a.k.a.~CCA -- \S\ref{s:G17m}).

The self-regulated, bipolar AGN outflows propagate outwards and gently dissipate the mechanical energy 
via bubble mixing, turbulence, and weak shocks, thus restoring most of the previously radiated internal energy (i.e., a global quasi thermal equilibrium),
with a duty cycle of the order of the central $t_{\rm cool}$ (\citealt{Gaspari:2017_uni} for a review). This simulation has been shown to be consistent with several spectroscopical X-ray constraints, including the suppression of the cooling flow by over 20 fold and the Gyr survival of the cool core (preserving the positive $T$ gradient; \citealt{Gaspari:2013_rev}).

\subsection{G17 simulation: ultra high-resolution AGN feeding} \label{s:G17m}

While G12 runs model a more realistic AGN feedback injection, the G17 (and previously related) simulations aim to resolve with maximally feasible resolution ($0.8\,$\;pc) the top-down multiphase condensation cascade, from the keV plasma phase to the {\it warm} gas ($10^3$\,-\,$10^5$\;K) and to the {\it cold} (20\,-\,200\;K) molecular gas. The simulation zooms in, with static mesh refinement, on the central massive galaxy (akin to NGC 5044) in the inner gaseous cool core (52$^3$ kpc$^3$ domain; again, minimum plasma $t_{\rm cool}/t_{\rm ff}< 10$), reaching a 100 Myr evolution. The static potential over such a small domain is dominated by the central galaxy stellar mass $M_\ast\simeq3.4\times10^{11}\;\msun$ (with effective radius $R_{\rm e}\simeq10$\;kpc).
The central SMBH ($M_\bullet = 3\times10^9\;\msun$) is modeled with the pseudo-relativistic \citet{Paczynski:1980} potential and has a characteristic Bondi radius of 85\;pc (G17, Sec.~2.2). 

The plasma radiative emissivity follows the same prescription as described in \S\ref{s:G12m} ($\simeq n^2\,\Lambda$)
though now the cooling curve is extended down to the neutral and molecular regime (Fig.~1 in G17) following an analytic prescription analogous to reference ISM studies (e.g., \citealt{Dalgarno:1972,Kim:2013}; Sec.~2.5 in G17). The included processes are atomic line cooling from hydrogen Ly$\alpha$, CII, OI, FeII, SiII; rovibrational line cooling from H$_2$ and CO; and atomic and molecular collisions with grains (\citealt{Koyama:2000}). The typical ionization fraction in the warm/cold phase is of the order of 1\% (mimicking the influence of post-AGB stars, AGN, and cosmic rays).

As we are here interested in the feeding stage and due to the limited integration time, we model only the gentle 4$\pi$ deposition stage of the mechanical AGN feedback, which prevents the cooling flow catastrophe and related monolithic collapse of the hot atmosphere (G12, Fig.~9). The injected plasma halo heating rate thus balances the average cooling rate in coarse radial shells (G17, Sec.~2.4), i.e., the hot atmosphere is globally stable but locally unstable.
We further include heating of the cold/warm phase (G17, Sec.~2.5), which is mainly due to the photoelectric effect; however, since massive/early-type galaxies have minor star formation ($\lta0.01\,\msun$\;yr$^{-1}$), this term is subdominant.
We note that we are fully aware of the complexity of the heating and cooling microprocesses in the multiphase ICM/IGrM; these simulations are part of our ongoing numerical campaign to dissect each physics step by step 
(\S\ref{s:lim}).

The additional key physics, alongside hydrodynamics, cooling, and heating, is turbulence.
As probed naturally in the previous self-regulated AGN outflow runs (and independent studies; \S\ref{s:intro}), the hot X-ray halo is continuously stirred by subsonic turbulence (due to the buoyant bubbles, Kelvin-Helmholtz instabilities, and shocks). 
Being these runs controlled experiments, the diffuse hard X-ray halo is continuously perturbed by subsonic turbulent motions ($\sigma_v\approx170\;\kms$) via a spectral forcing scheme based on an Ornstein-Uhlenbeck random process (G17, Sec.~2.3), which also reproduces experimental high-order structure functions (\citealt{Fisher:2008}). The gas is only stirred at low-$k$ Fourier modes,
allowing the development of a self-consistent and natural turbulence cascade. Such a subsonic, mainly solenoidal turbulence mimics well that produced by AGN feedback during the deposition stage (e.g., \citealt{Gaspari:2012b}).
Since we focus in this work on massive galaxies, the driven chaotic gas motions are stronger than the gas rotational velocity (i.e., ${\rm Ta} \equiv v_{\rm rot}/\sigma_v < 1$). 

In this heated and turbulent atmosphere, warm filaments and cold clouds condense out of the hot plasma, some of which rain on the SMBH.
In the nuclear region inelastic collisions allow the angular momentum to mix and cancel via chaotic cold accretion. As shown in our previous series of investigations (cf.~\citealt{Gaspari:2013_cca,Gaspari:2015_cca,Gaspari:2017_cca}), CCA displays important properties that can explain diverse observed phenomena, including the rapid flickering of AGN, their obscuration properties (the broad-/narrow-line region and clumpy torus), the shallow X-ray temperature profiles, and the cospatiality of the inflowing/outflowing multiphase gas in the soft X-ray, optical, and radio bands. In this work, we focus on the multiphase CCA kinematics.\\

\section{Linking the multiphase gas kinematics} \label{s:res}
\noindent
While in previous works we focused on the thermodynamics and accretion process, 
here we analyze the LOS luminosity-weighted (LW) 
kinematics of the multiphase gas and possible correlations. Numerically, the mean LW LOS velocity is computed as
\begin{equation} \label{e:vlos}
\bar{v}_{\rm los} = \frac{\Sigma_{\rm los}\,v_k\,\Delta L_k}{\Sigma_{\rm los}\,\Delta L_k}, 
\end{equation}
and the related velocity dispersion $\sigma_{v,{\rm los}}$ as
\begin{equation} \label{e:sigmav}
\sigma_{v, {\rm los}}^2 = \frac{\Sigma_{\rm los} (v_k - \bar{v}_{\rm los})^2\,\Delta L_k}{\Sigma_{\rm los}\,\Delta L_k}, 
\end{equation}
where the summation is computed over a given cylindrical aperture along the full line of sight (with the SMBH as center).
The luminosity for each cell $k$ with volume $\Delta V_k$ is $\Delta L_k \simeq n_k^2\Lambda_k\,\Delta V_k$, 
where $\Lambda$ is the radiative cooling curve (\S\ref{s:G17m})
in the temperature band of the given gas phase.
In the following subsections, we mainly analyze the numerical results, while we dedicate \S\ref{s:obs} to an in-depth comparison with recent and new observations of multiphase gas in massive galaxies.\\

\begin{figure}
\centering
\subfigure{\includegraphics[width=1.01\columnwidth]{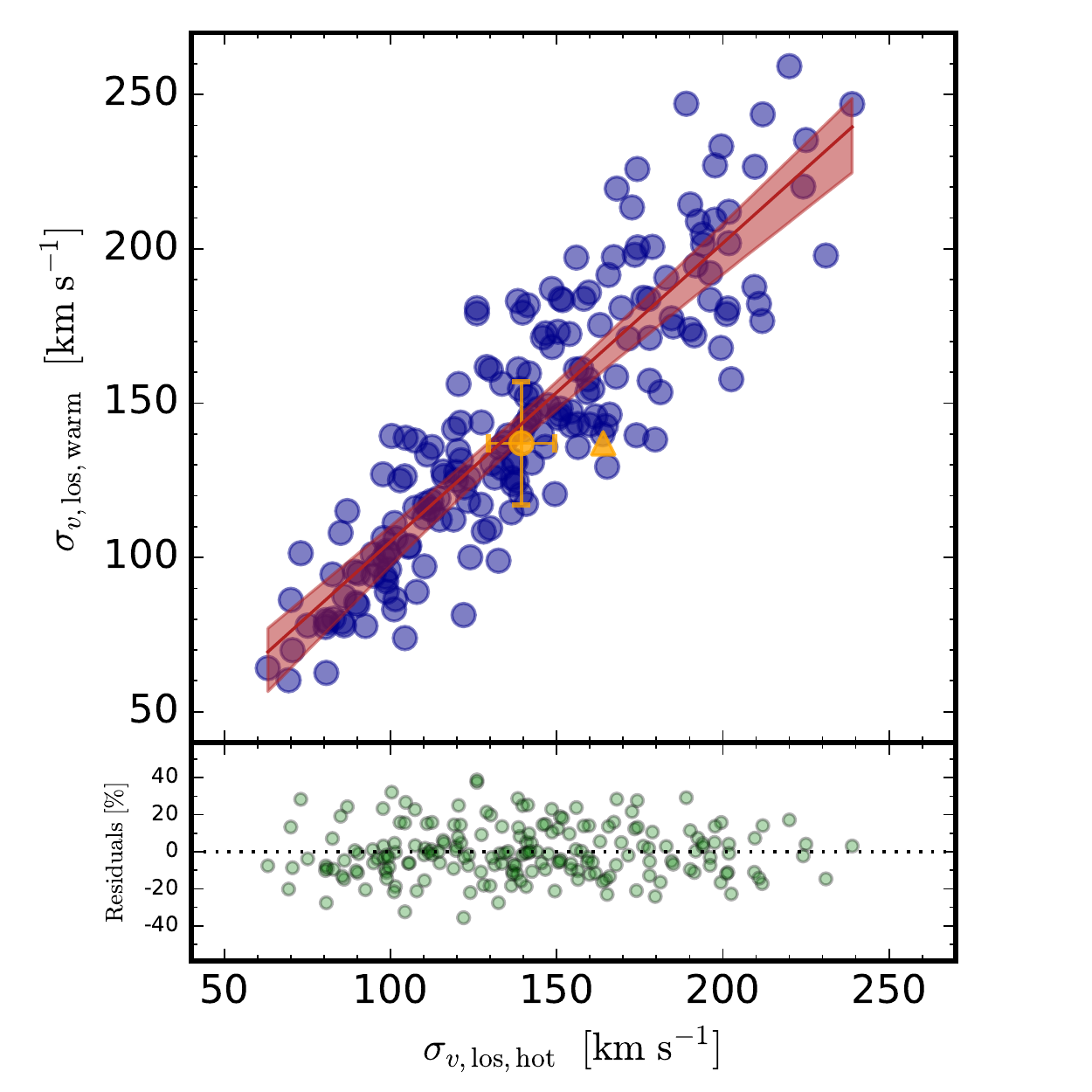}}
\caption{Long-term self-regulated mechanical AGN feedback simulation in a typical galaxy cluster core. {\it Top:} Correlation between the luminosity-weighted LOS velocity dispersion of the ensemble warm gas ($5\times10^3$\,-\,$5\times10^4$\;K) and X-ray plasma ($0.3$\,-\,8\;keV) 
in the core region ($4\le R/{\rm kpc} \le 45$), plotted every 10 Myr for over 2 Gyr. 
Each point is the median value from five random lines of sight at each time. 
The brown line and bands show the best-fit linear regression (Pearson $r=0.87$) and the associated 99\% confidence levels retrieved via bootstrap resampling with $10^5$ iterations. 
The best-fitting values for the slope and normalization are $0.97^{+0.01}_{-0.02}$ and $8.3^{+3.5}_{-5.1}\ \kms$, respectively. 
The orange points show the observational constraints from the Perseus cluster combining the SITELLE H$\alpha$+[NII] data with the {\it Hitomi} $\sigma_{v, {\rm los, hot}}$ detection.
We plot the {\it Hitomi} iron-lines spectroscopic measurements obtained by fitting the hard X-ray band (orange triangle). As simulated hard X-ray plasma velocity dispersions are on average 20\% higher than those in the whole X-ray band, we plot the {\it Hitomi} measurement decreased by this amount (orange circle).
{\it Bottom}: Percent residuals of the simulated points from the best-fit relation. 
The {\it ensemble} warm phase kinematics behaves as a quasi linear tracer of the diffuse plasma turbulence, so it is possible to reliably convert between the two velocity dispersions, taking advantage of the low-energy bands (e.g., optical/IR).
\\
} 
\label{f:G12_2corr}
\end{figure}

\vspace{+0.15cm}
\subsection{Long-term self-regulated kinetic AGN feedback} \label{s:G12}

\noindent
We start from the long-term simulation, 
which can follow condensation only down to the warm phase but in a long-term AGN outflow feedback evolution (G12; \S\ref{s:sims}). Fig.~\ref{f:G12_2corr} addresses a key question: what is the degree of correlation between the velocity dispersions (spectral line width) of the condensed {\it ensemble} warm gas and the volume-filling X-ray plasma in cluster cores?

During the Gyr evolution, the hot halo is continuously perturbed by the cosmic weather and the AGN outflows at large and small radii, respectively. The turbulent motions promote the nonlinear condensation of extended warm ionized filaments ($T\approx10^4$\;K), which are mainly observed in H$\alpha$+[NII] emission.
Fig.~\ref{f:G12_2corr} shows that during the top-down multiphase condensation and recurrent AGN feedback cycles (blue points),
the ensemble warm phase behaves as a quasi linear tracer of the turbulent eddy evolution (see also \S\ref{s:bro}).
The best-fit linear relation has the following slope and normalization:
\begin{equation}\label{e:fit}
\sigma_{v,{\rm los, warm}} = 0.97^{+0.01}_{-0.02}\;\sigma_{v,{\rm los, hot}} + 8.3^{+3.5}_{-5.1}\ \:\kms.
\end{equation}
The linear correlation emerges within a wide ensemble extraction region of size 1-2 condensation radii 
(for our massive cluster $\sim$\,45\;kpc),
over which the warm gas forms out of the progenitor X-ray (0.3\,-\,8\;keV) plasma. 
At small scales, the single clouds show instead a larger variance driven by the local eddies and cloud-collisional kinematics (see \S\ref{s:G17}).

The turbulence eddy turnover timescale tied to this coupling region (which is also comparable to the AGN bubble injection scale) is the effective dynamical timescale of the top-down
multiphase condensation process (\S\ref{s:crit}). 
For well-resolved objects, it is preferable -- but not essential -- to cut the very inner region (here 4 kpc) in the presence of jets activity\footnote{As long as the wide aperture captures the bulk of the condensed gas and related emission, the ensemble detection is not sensitive to changing the inner/outer extraction radius by a few kpc, remaining within the retrieved scatter.}.
The simulated velocity dispersion distribution has mean $\sigma_{v,{\rm los}}\approx 140\;\kms$,
reaching values up to 250$\;\kms$ during the stronger AGN feedback phases. 
The logarithmic scatter of the entire warm/hot gas distribution is 0.13 dex.
Focusing instead on the deviation from the best-fit line,
the RMS is 14\% (with maximum residuals up to $\pm40$\%) which is mainly generated by the AGN duty cycle and related time hysteresis between driving perturbations and recurrent residual condensation. 

In \S\ref{s:obs}, we compare the simulated distribution with new ensemble warm gas constraints for 76 clusters, resulting to be consistent with the simulated range predicted here.
For one cluster -- Perseus -- we can directly probe the correlation here, as direct LOS velocity dispersion detections for both the warm gas and X-ray plasma are available.
Specifically, we combine the Fe\,XXV-XXVI linewidth fiducial detection, $\sigma_{v,{\rm los, hot}}=164\pm10\;\kms$ (\citealt{Hitomi:2016}\footnote{Recently, a few more uncertain regions have been added to the analysis, which nevertheless resulted in a similar single-spectrum value $\sigma_{v,{\rm los, hot}}=153^{+21}_{-27}\;\kms$(\citealt{Hitomi:2017}, Sec.~3.4).}) , with a new wide-field SITELLE\footnote{A wide-field imaging Fourier transform spectrometer (IFTS) with IFU capabilities in the visible (350\,-\,900 nm) for the Canada-France-Hawaii telescope (CFHT; \citealt{Drissen:2010}): \href{http://www.cfht.hawaii.edu/Instruments/Sitelle}{http://www.cfht.hawaii.edu/Instruments/Sitelle}.} IFTS observation\footnote{Data taken in January 2016 with the SN3 filter (651-685 nm) for 2.14h (308 exposures of 25\,s), with a spectral resolution of 1800. The data reduction and calibration were conducted by using ORBS and the analysis tools from ORCS (\citealt{Martin:2015}).}
of the ensemble H$\alpha$+[NII] linewidth (Gendron-Marsolais et al.~in prep.; see \S\ref{s:obs_ens} for the analysis method).
By fitting the H$\alpha$+[NII] lines of the single spectrum integrated over the same wide extraction region as above,
we find $\sigma_{v,{\rm los,warm}}=137\pm20\;\kms$.  
Selecting the hard X-ray band as for the {\it Hitomi} iron-lines measurements ($\approx$\,5\,-\,9\;keV), we find simulated plasma velocity dispersions that are on average 20\% higher than those in the entire X-ray band, since the hard X-ray, less dense gas is more easily accelerated by feedback. Nevertheless, whether or not this correction is applied (orange circle versus triangle in Fig.~\ref{f:G12_2corr}), the warm and hot gas velocity dispersions are consistent with having comparable turbulent kinematics within uncertainties, 
in agreement with the predicted correlation.

\vspace{+0.15cm}
\subsection{High-resolution chaotic cold accretion feeding} \label{s:G17}
\noindent
We move on from the long-term evolution to the detailed ultra high-resolution kinematics of the top-down multiphase condensation (0.8\;pc -- ensuring convergence of the main properties), which tracks all the phases down to the molecular regime (G17 and \S\ref{s:sims}). While the previous simulation focuses on the realistic feedback process, the current run focuses on the detailed feeding process in a central massive galaxy for a shorter time, 100 Myr, which is still 10 times the central (kpc-scale) cooling time. In this turbulent and heated halo, extended warm filaments ($\sim10^4$\,-\,$10^5$\;K) condense out of the hot keV plasma atmosphere and produce a condensation rain. 
The thin outer layer of the filament is ionized and strongly emitting in optical and UV. The spine of the filaments is mostly neutral gas ($\sim\,$$10^3$\;K), containing most of the warm gas mass. The denser peaks further condense into molecular clouds ($< 100$\;K) with radii spanning several pc to 100 pc for the giant molecular associations. Total molecular masses can reach up to several $10^7\;\msun$, consistent with recent ALMA data (e.g., \citealt{David:2014}; massive clusters can show even $10^9\;\msun$, e.g., \citealt{Vantyghem:2016,Pulido:2017}).
While temperature radial profiles are fairly flat for all condensed phases, density profiles have logarithmic slope -1, with inner densities up to $10^{-21}$\;g\,cm$^{-3}$ for the molecular phase.

The CCA process is also responsible for efficiently boosting SMBH accretion rates with rapid intermittency up to two orders of magnitude (e.g., \citealt{Gaspari:2016} for a brief review).
Given that subsonic turbulence is a common state of hot halos (\S\ref{s:obs}), CCA is also a recurrent state of observed systems (e.g., \citealt{McDonald:2011a,McDonald:2012_Ha,Werner:2014,Tremblay:2016,David:2017}), although overheated halos can experience a pure hot low-accretion mode, and rotation-dominated halos can be associated with a decoupled thin disk.
We refer the interested reader to G17 for the detailed thermodynamic properties of each phase and in-depth discussions; here, are interested in the statistical kinematic properties related to the CCA rain, in particular confronting the ensemble versus local variance and the mean of the velocity field (i.e., the broadening and shift of the spectral lines) for all the gas phases. The limitations of the current simulations and future improvements are discussed in \S\ref{s:lim}.

\vspace{+0.15cm}
\subsubsection{Turbulence: line broadening} \label{s:bro}
\begin{figure}
\centering
\hspace{-0.15cm}
\subfigure{\includegraphics[width=0.90\columnwidth]{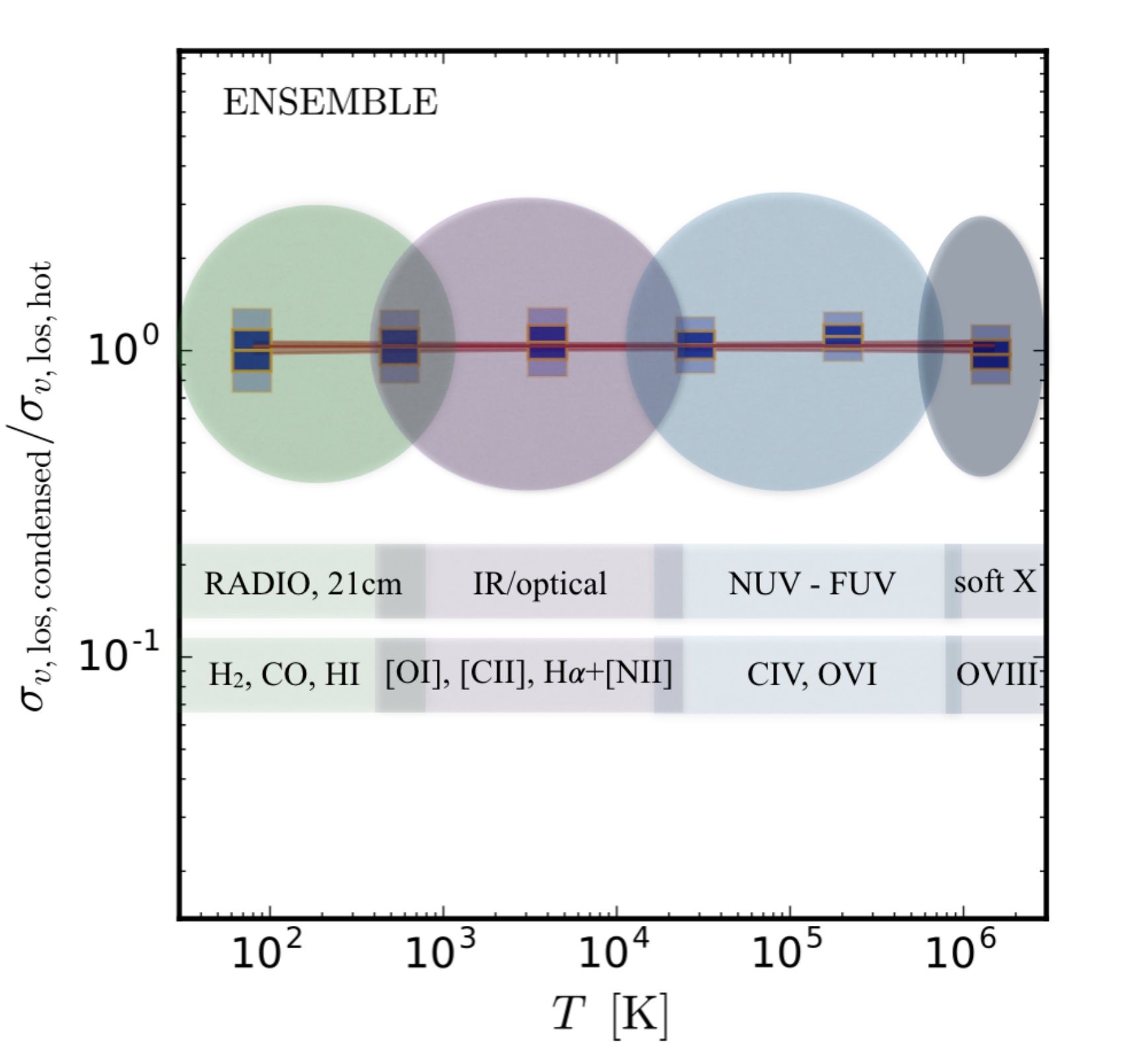}}
\subfigure{\includegraphics[width=0.913\columnwidth]{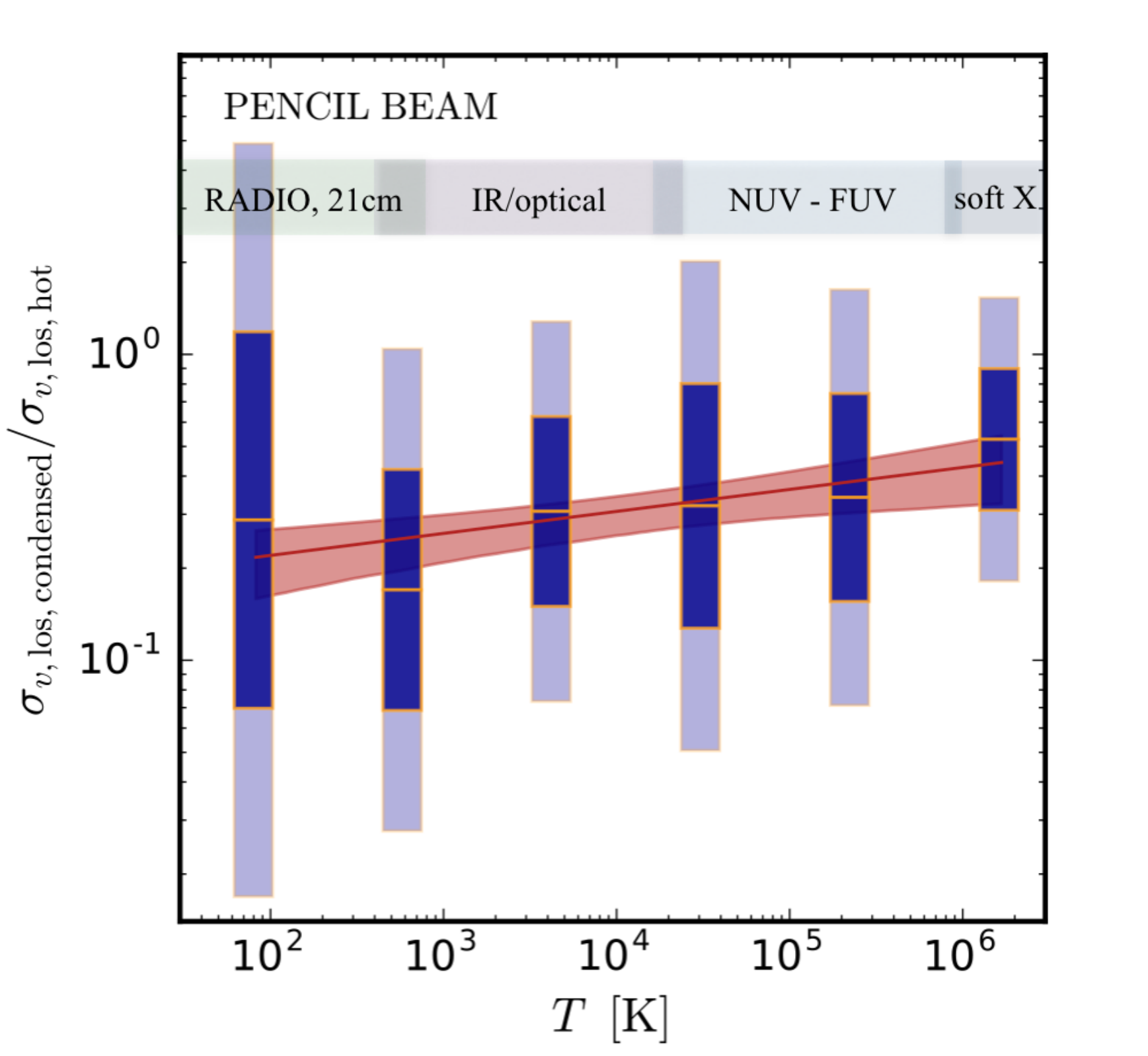}}
\caption{Ultra high-resolution (0.8\;pc) simulation following the multiphase CCA rain for 100\;Myr in a central massive galaxy:
luminosity-weighted LOS velocity dispersion (line broadening) of the multiphase gas (from ionized to molecular gas, 6 bins with 0.9 dex width) normalized to the hot plasma ($> 5\times10^6$\;K) turbulence. 
The latter (not shown) is continuously driven with a stable one-dimensional velocity of the order of $100\;\kms$.
The bars indicate the mean and 1-2 (dark to light) standard deviations in log space of the underlying points, which are tracked every 1 Myr.
Each point is computed as the median out of 5 random lines of sight.
We note, for a Gaussian line, ${\rm FWHM} = 2\sqrt{2\ln2}\,\sigma_{v,{\rm los}}$. 
The brown line and bands show the best-fit linear regression (in log space)
and associated 99\% confidence levels (with $10^5$ iteration bootstrapping) for the computed points over the 6 phases. 
{\it Top:} Ensemble-beam detection in the projected radial range 0.5\,-\,15\;kpc.
The correlation is tight throughout the multiphase condensation cascade, 
thus we can use the ensemble detection as robust proxy for the turbulence in the hot gas (or between other phases). 
{\it Bottom}: Same as above but for a pencil-beam ($R\lesssim$\,25\;pc) observation. 
The velocity dispersion decreases significantly, in particular for the cold phase, 
thus narrow lines are expected to be detected frequently with this technique.
The scatter increases substantially,
implying that systems observed with a pencil beam (e.g., through the AGN backlight or CO emission) can also display a broad component. The narrower component is typically associated with inner denser clouds, which have experienced several collisions and are being funneled toward the supermassive black hole.
} 
\label{f:G17_broad}
\end{figure}

\noindent
Fig.~\ref{f:G17_broad} shows the ensemble velocity dispersion in six major temperature bins normalized to the subsonic turbulent velocity, which is stably driven in the hot plasma ($>5\times10^6$ K). 
The different phases correspond to key observational bands, covering the radio, optical/IR, and UV/soft X-ray regimes, as highlighted by different colors in the top panel.
Turbulent LOS velocity dispersions are detected through the broadening of the observed spectral lines by measuring the line's full width at half maximum, ${\rm FWHM}\simeq2.355\,\sigma_{v,{\rm los}}$. The lower the temperature, the smaller the contribution of thermal broadening\footnote{The relative turbulent and thermal Doppler broadening are respectively given by $\Delta\nu_{\rm turb}/\nu_0=\sigma_{v,{\rm los}}/c$ and $\Delta\nu_{\rm th}/\nu_0=(2k_{\rm b}T/m_{\rm i})^{1/2}/c$, where $\nu_0$ is the line center frequency and $m_{\rm i}$ is the mass of the given ion.}, which is $\sim$\,1\,-\,8\;$\kms$ for molecular and warm gas, respectively.
The H$_2$, CO, HI, [CII], and H$\alpha$+[NII] lines are all excellent probes of the gas kinematics.

The top and bottom panels of Fig.~\ref{f:G17_broad} show the same velocity dispersion diagnostics for the ensemble-beam (excluding the collisional nuclear region) and for a pencil-beam (aperture $R\lta 25$ pc) detection, respectively.
The blue bars indicate the logarithmic mean and 1-2 standard deviation\footnote{
The average fractional difference (as absolute value) between the mean/1$\sigma$ and median/$\pm$34.1\% interval for the ensemble-beam points in log space is 7.8/8.6\%, respectively. Gaussian fits are thus a good representation of the distributions and are accurate enough for the scope of the present work.} 
of the underlying points (not shown), which are tracked every 1 Myr for $\sim$100 Myr.
The ensemble measurement substantially reduces the turbulence intermittency noise and shows again a tight linear correlation throughout the phases, corroborating the result in \S\ref{s:G12}.
The RMS deviation from the hot gas turbulent velocity is 13\% (brown), which is analogous to the long-term simulation deviation from the best-fit line in Fig.~\ref{f:G12_2corr}.
The ratio is not unity as condensed structures do not fill the entire halo at every moment in time.
This demonstrates that we can use the ensemble warm or cold gas as tracers of the kinematics of the turbulent hot gas, and vice versa we can predict the kinematics of the multiphase CCA cascade from the turbulent plasma halo.
The optical/NUV phase near the stable $10^4$\;K regime has one of the lowest scatters and better equivalence, making nebular H$\alpha$+[NII] emission an excellent tool to study turbulence (\S\ref{s:obs_ens}).
The FUV phase shows larger mean velocity dispersion, experiencing the most rapid and unstable condensation transition due to the strong line cooling, while tracing the low-mass filament skin (cf.~G17 for the multiwavelength synthetic imaging). The molecular clouds, having the lowest volume filling, display the largest scatter.
Globally, the condensed gas cannot be treated as ballistic or free-falling objects, as all phases participate in the hydrodynamical layer-within-layer cascade. Note that although some of the condensed gas can be accreted by the SMBH, the phases are continuously replenished by the turbulent condensation rain.

The bottom panel of Fig.~\ref{f:G17_broad} shows that, in
the synthetic observations with a pencil beam (small aperture through the center),
the velocity dispersion decreases significantly, down to a few 10\% of the hot gas value.
Therefore, we expect to detect commonly narrow lines with this approach, with FWHM down to a few 10 $\kms$.
At the same time, the scatter increases substantially (the distribution is lognormal with dispersion over all the phases of 0.41\;dex), 
thus a broad component can also be present in pencil-beam measurements (e.g., by using absorption lines against the AGN backlight; \S\ref{s:obs_pen}).
The broad component is typically associated with structures at large radial distance having small line shift. 
The narrower component tends instead to track the inner denser clouds (especially for the colder phases), which have experienced inelastic collisions in the nuclear region and are being funneled toward the SMBH.
Such clouds can be better probed via major blue-/redshifted lines (\S\ref{s:shift}). 

The increased scatter (bottom versus top panel) reflects the chaotic and intermittent nature of turbulence, since the single warm/cold structures fill smaller volumes while condensing down the turbulence cascade. 
The pencil-beam approach indeed tends to sample a few or single clouds (e.g., for the molecular phase, the inner volume filling is 2\%, gradually decreasing beyond $r>1\;$kpc).
Because of the turbulence inertial cascade, the velocity dispersion decreases\footnote{By using the power spectrum analysis tool developed in \citet{Eckert:2017_PS}, we checked that in projection the Kolmogorov cascade retains the same power index, in particular at small scales.} as $\sigma_v\propto l^{1/3}$. From characteristic 2\;kpc to 20\;pc scales, this implies a factor of 0.2 decrease in velocity dispersion, as retrieved for the molecular phase in Fig.~\ref{f:G17_broad}. Warmer phases are less compressed, thus having a lower decrement, as shown by the positive best-fit line slope.

The scatter related to multiple off-center pencil-beam measurements of the condensed gas line broadening can be used as a new way to quantify the level of small-scale intermittency in the turbulent medium.
It is worth noting that the cold molecular phase suffers the largest scatter in line broadening.
While the numerous cold clouds trace the condensation out of the peaks of the filamentary warm gas (in turn formed out of the turbulent hot halo), they also experience chaotic collisions and drag with all other phases, in particular at small distances from the SMBH.
A fraction of the clouds may turn into young star clusters, decoupling via the collisionless dynamics (likely retaining the progenitor cold gas velocity dispersion). However, a significant $\sigma_v$ in all the condensed gas phases implies that turbulent pressure is a key component (dominating over thermal pressure) which prevents most clouds from major collapse (cf.~G17), in agreement with the low mean star formation rates and large cloud virial parameters ($\alpha \gg 1$) observed in early-type galaxies (e.g., \citealt{David:2014,Temi:2017}).\\

\vspace{+0.15cm}
\subsubsection{Bulk/inflow motions: line shift} \label{s:shift}
\noindent

\begin{figure}
\centering
\subfigure{\includegraphics[width=0.910\columnwidth]{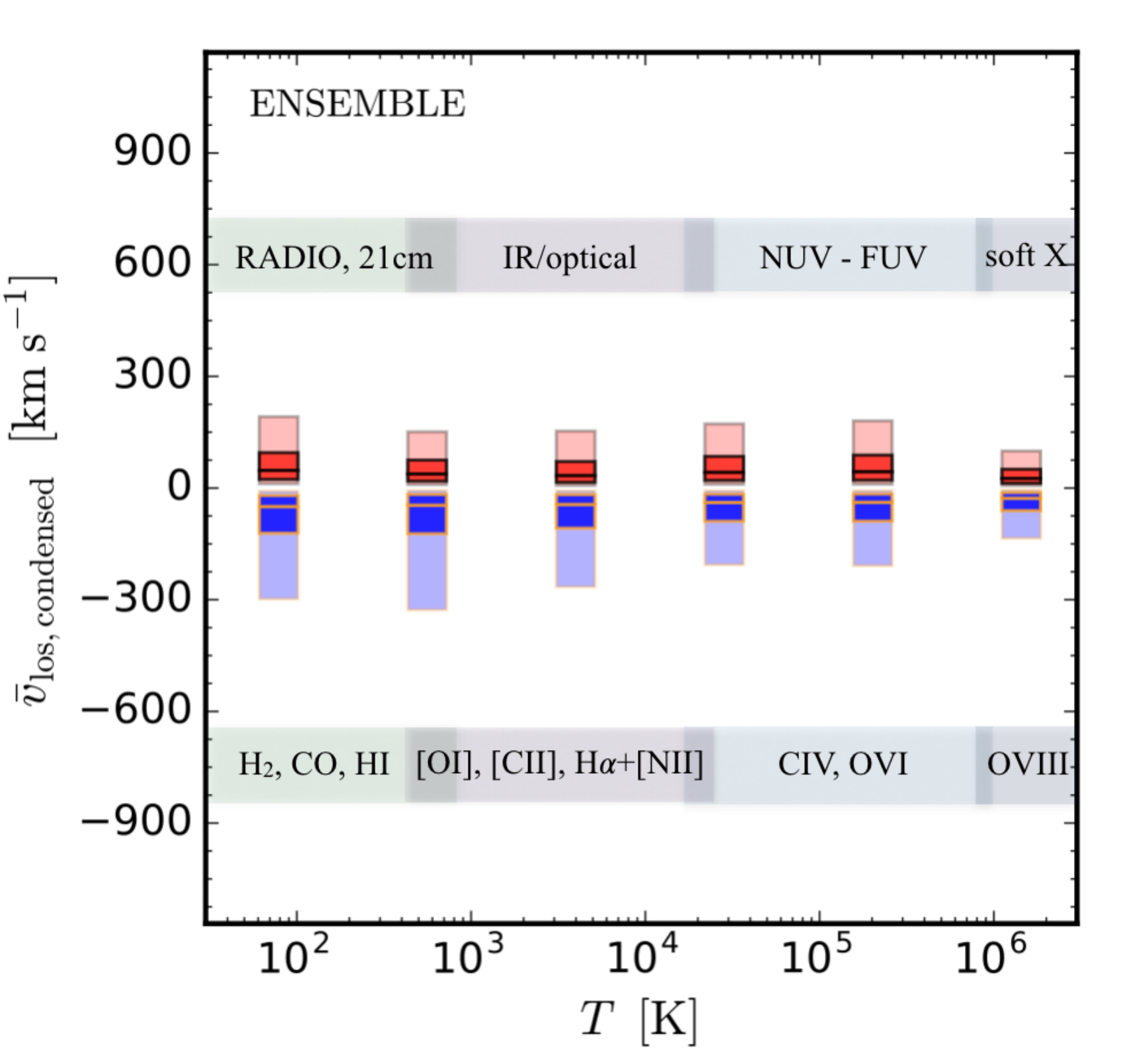}}
\subfigure{\includegraphics[width=0.898\columnwidth]{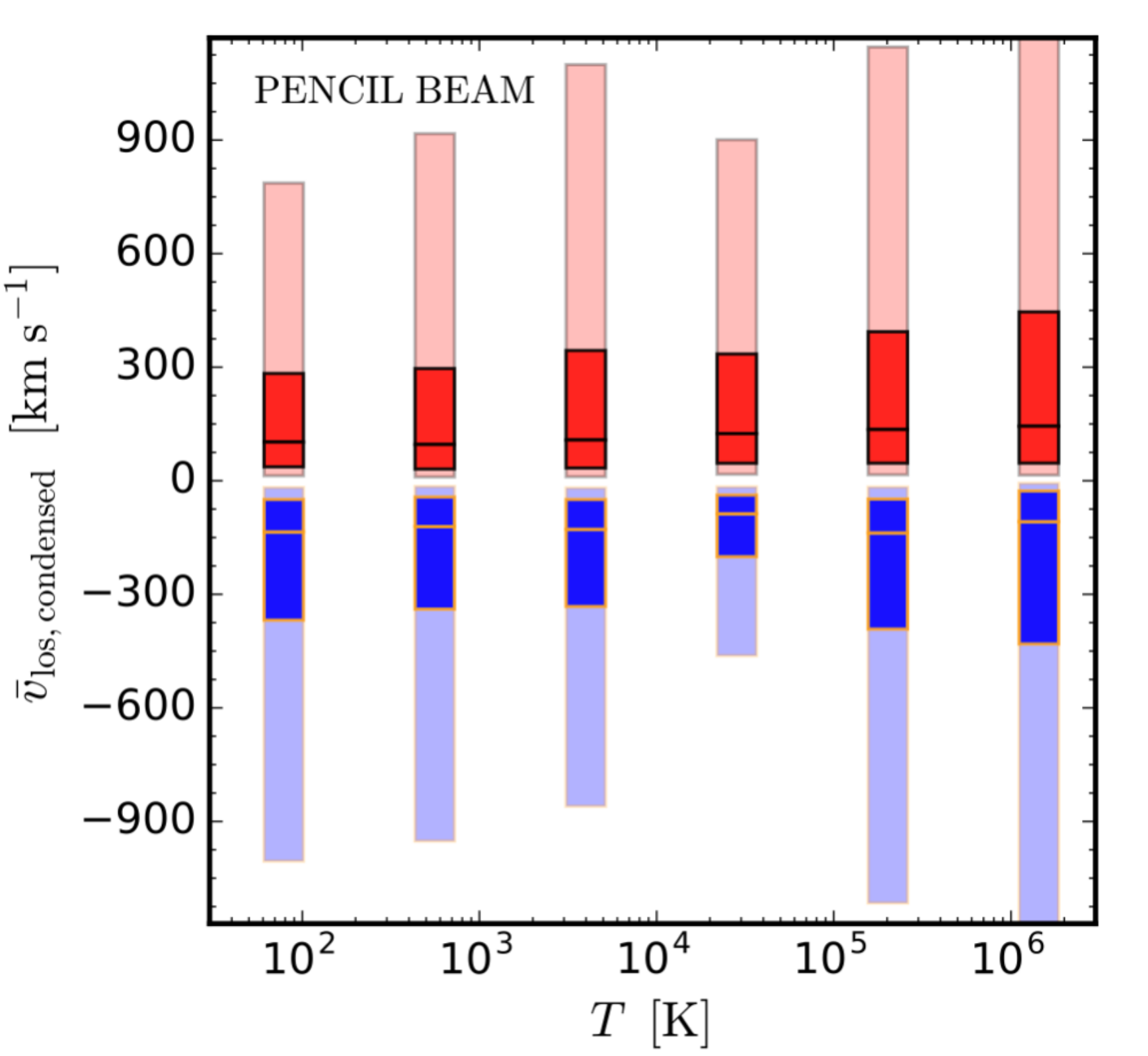}}
\caption{Ultra high-resolution (0.8\;pc) simulation following the multiphase CCA rain for 100\;Myr in a central massive galaxy (analogue of Fig.~\ref{f:G17_broad}): luminosity-weighted LOS velocity (line shift) of the multiphase gas for the 6 phase bins.
The bars indicate the mean and 1-2 standard deviation (dark to light) in log space of the underlying points, which are tracked every 1 Myr for each of 3 random lines of sight. 
We note the driven hot keV plasma (not shown) has a null velocity shift with negligible error. Blue/red colors denote blue-/redshifted lines compared with the galactic systemic velocity. The line shift magnitudes are best fitted by lognormal distributions.
{\it Top:} The ensemble gas detections show on average a small line shift, with average magnitude less than a few tens $\kms$, and thus would sometimes appear consistent with the galactic dynamics. 
{\it Bottom:} The pencil-beam detections, on the other hand, show substantially larger line shifts of the order of several 100\;$\kms$, with a large scatter ($\sim0.5$\;dex). 
\\
}
\label{f:G17_shift}
\end{figure}

In Fig.~\ref{f:G17_shift}, we analyze the bulk motions during the same CCA run via the mean velocity along the line of sight, or analogously, via the spectral line blue-/redshift (as offset from the systemic velocity). We note that our galaxy (stellar) systemic velocity is always null, as the simulation box is centered on the (static) gravitational potential. 
The ensemble gas detection (top panel) typically shows a small line shift with fairly contained scatter, slightly increasing toward the cooler phase. The logarithmic mean and dispersion of its magnitude over all the phases are $\log |\bar{v}_{\rm los}|/(\kms)=1.59\pm0.37$. 
The line centroid would sometimes appear consistent with the galactic dynamics, given the typical measurement uncertainties. 
The pencil-beam detections (bottom panel) show instead a substantially larger line shift with global logarithmic mean and dispersion, 
$\log |\bar{v}_{\rm los}|/(\kms)=2.07\pm0.47$.
The fastest structures are typically inner clouds which have collided, canceling angular momentum, and are often falling toward the SMBH, within the Bondi capture radius.
In both cases, we expect on average a similar fraction of blue- and redshifted lines (at least in emission), as clouds can drift in front of or behind the accretor.\footnote{In absorption, a powerful outflow (on which some AGN surveys are selected) may skew the line absorption distribution toward the strongly blueshifted ($|\bar{v}_{\rm los}|> 500\;\kms$) side, swamping the infalling clouds. Conversely, during major angular momentum cancellation episodes, absorbed redshifted lines may be more frequent.
In observations, separating inflow from outflow is non-trivial and multiple constraints are essential (e.g., reverberation mapping). 
} 

Interestingly, the condensed gas kinematics distributions are best fitted by lognormal distributions.\footnote{
The average fractional difference (as absolute value) between the mean/1$\sigma$ and median/$\pm$34.1\% interval for the pencil-beam points in log space is just 2.6/7.5\%, respectively.}
This is particularly evident when the range increases to several dex as for the velocity shift, since the linear approximation is no longer valid, and the high-end tail of the distribution becomes prominent. The reason is that turbulence continuously drives nonlinear perturbations in all thermodynamic properties with a characteristic lognormal shape (e.g., \citealt{Gaspari:2014_coma2}). The turbulence cascade (and related multiphase condensation) is indeed a multiplicative process, with smaller and smaller eddies generated within larger vortices, which can also be seen as a power-law inertial range in Fourier space.

Overall, as shown for the line-broadening kinematics, the adoption of a pencil beam leads to sampling smaller structures, thus tracing a lower velocity dispersion and a larger velocity shift of the infalling multiphase clouds and filaments.
This method of probing the inner CCA via observations of narrow and significantly shifted lines is a simple and promising method, which is particularly efficient in absorption against a strong AGN backlight emitting in the band of the targeted gas phase (e.g., GHz radio for CO gas). An excellent case study is A2597 (\citealt{Tremblay:2016}), where ALMA detected 3 central infalling narrow-line clouds with redshifted velocities up to $335\;\kms$ (\S\ref{s:obs_pen} for a large dataset comparison).\\

\subsection{Additional physics} \label{s:lim}
The previous simulations are part of our numerical campaign to dissect the multiphase physics of clusters, groups, and massive galaxies, as we test and disentangle each physical process step by step.
We discuss below the main limitations of the current runs.
In general, such extra -- typically subdominant -- physics tends to alter the details of the condensed gas (morphology, ionization layers, etc.), while the statistical thermodynamic and kinematic properties are expected to remain similar, i.e., the gas condenses via the turbulent top-down cascade, feeding the SMBH via accretion of clouds and filaments that are disordered on large scales.

Magnetic fields and cosmic rays (CRs) can provide non-thermal pressure, in addition to turbulence, and further support the condensation collapse, thus altering the size of the clouds and filaments (e.g., \citealt{Sharma:2010}). The draping provided by the large-scale $B$-fields around the warm and cold structures is also expected to mitigate the heat and mass exchange between the different phases. On the other hand, the typical strength of magnetic fields is a few $\mu$G, thus $\beta \gg 1$ in the hot phase, implying that they are dynamically unimportant over most of the volume. Regarding CRs, their transport properties (as diffusion and streaming) out of the Galaxy are highly uncertain. {\it Fermi} telescope has also put severe upper limits on the amount of gamma-ray emission in the ICM, with no significant signal even in stacked analyses (e.g., \citealt{Huber:2013}), implying CR-to-thermal energy ratios of less than a few percent. 

Speaking of diffusion processes, anisotropic conduction and viscosity will likely promote the formation of more elongated filamentary structures, with more equilibrated temperature and momentum along the magnetic lines. However, ram-pressure stripping observations (e.g., \citealt{DeGrandi:2016,Eckert:2017_ramP}), analytic studies (e.g., \citealt{Burkert:2000}), and plasma particle-in-cell simulations (e.g., \citealt{Komarov:2016,Kumar:2017}) suggest that the transport Spitzer/Braginskii coefficients are suppressed by at least one to two orders of magnitude due to plasma micro-instabilities (e.g., firehose and mirror) and/or highly tangled magnetic fields below the Coulomb mean free path.

Hot stars and AGN radiation can all contribute to ionize the external layers of the warm filaments. 
While the ionized skin depth varies widely depending on the clump density and location 
(cf.~\citealt{Valentini:2015}), the typical heat input is modest in massive/early-type galaxies (see \citealt{Loewenstein:1990}), which have both very low star formation rates and Eddington ratios.
Connected to this, stellar feedback is several orders of magnitude lower compared with AGN heating (e.g., \citealt{Gaspari:2012b}). Self-gravity is also negligible as most of the clouds have a large virial parameter because of non-thermal pressure, as found in ALMA data (e.g., \citealt{David:2014,Temi:2017}). 
Despite its simplicity, our feeding runs showed that radiative emission from line recombination during the condensation cascade down to $10^4$\;K, together with mixing with the hot plasma, can reproduce H$\alpha$ maps similar to those retrieved by SOAR (\citealt{Gaspari:2015_cca}, Sec.~8). Regardless of the ionization level and microphysics, the turbulence cascade is a top-down process that develops in an analogous way, generating extended filaments and cloud associations that trace the self-similar turbulent eddies. 

Overall, although more physics is likely at play in the ICM/IGrM -- in addition to gas dynamics, gravity, radiative multiphase cooling, AGN heating, and turbulence implemented here -- the current simulations already provide a robust framework and laboratory to assess the main statistical properties of the multiphase condensation cascade. 
Passing different multifrequency observational tests (e.g., radial profiles, surface brightness maps, cooling rates, emission measures, AGN variability; cf.~G12-G17) gives confidence that the simulations are already capturing the main features of the real condensation process. 
Our forthcoming works will be important  to dissect in depth the above physics and assess any deviation from the current results, thus improving the theoretical understanding of the multiphase gas precipitation.
\\

\capstartfalse
\begin{center}
\begin{deluxetable*}{lllclllc} 
\setlength{\tabcolsep}{2pt}
\tabletypesize{\small}
\tablecaption{Detected line broadening and shift for the {\it ensemble} warm and cold phase in observed massive galaxies within cluster/group cores. \\ 
These constraints can be used by other studies as proxies of the turbulent velocities and/or bulk motions of the diffuse ICM/IGrM.
}
\tablehead{
\colhead{Object\ \ \ \ } & \colhead{$\sigma_{v,{\rm los}}$ [$\kms$]} & \colhead{$\bar{v}_{\rm los}$ [$\kms$]} & \colhead{$R_{\rm ex}$\,[kpc]} & \colhead{\vline\ Object\ \ \ \ \ } & \colhead{$\sigma_{v,{\rm los}}$ [$\kms$]} & \colhead{$\bar{v}_{\rm los}$ [$\kms$]} & \colhead{$R_{\rm ex}$\,[kpc]}
}
\startdata
 A1348\tablenotemark{a}   &   H$\alpha$+[NII]: $281\pm3$  & H$\alpha$+[NII]: $-76\pm8$  & 18.0 & 
\vline \  A1663\tablenotemark{a}   &    H$\alpha$+[NII]: $241\pm3$  & H$\alpha$+[NII]: $34\pm9$ & 13.7  \\
 A1060\tablenotemark{a}  &     H$\alpha$+[NII]: $90\pm2$  & H$\alpha$+[NII]: $-54\pm14$  & 1.2 & 
\vline \  A133\tablenotemark{a}    &   H$\alpha$+[NII]: $122\pm2$  & H$\alpha$+[NII]: $0\pm14$ & 7.2   \\
 A1668\tablenotemark{a}   &   H$\alpha$+[NII]: $166\pm3$  & H$\alpha$+[NII]: $-72\pm10$  &  10.9  &
\vline \  A2052\tablenotemark{a}   &   H$\alpha$+[NII]: $162\pm4$  & H$\alpha$+[NII]: $-31\pm11$  & 6.4     \\
 A2415\tablenotemark{a}   &   H$\alpha$+[NII]: $170\pm3$  & H$\alpha$+[NII]: $-85\pm12$  &   11.2    &
\vline \  A2495\tablenotemark{a}   &   H$\alpha$+[NII]: $128\pm4$  & H$\alpha$+[NII]: $0\pm11$  & 8.9   \\
 A2566\tablenotemark{a}   &    H$\alpha$+[NII]: $149\pm3$  & H$\alpha$+[NII]: $12\pm10$  &   10.1    &
\vline \  A2580\tablenotemark{a}   &    H$\alpha$+[NII]: $85\pm5$  & H$\alpha$+[NII]: $-44\pm14$  &  9.4  \\
 A2734\tablenotemark{a}   &   H$\alpha$+[NII]: $229\pm3$  & H$\alpha$+[NII]: $-42\pm11$  &  6.1     &
\vline \  A3112\tablenotemark{a}    &  H$\alpha$+[NII]: $258\pm3$  & H$\alpha$+[NII]: $0\pm11$ &  8.9   \\
 A1084\tablenotemark{a}   &     H$\alpha$+[NII]: $135\pm3$  & H$\alpha$+[NII]: $59\pm12$  & 11.2      &
\vline \  A3581\tablenotemark{a}   &   H$\alpha$+[NII]: $188\pm3$  & H$\alpha$+[NII]: $-19\pm11$ &  5.4   \\
 A3605\tablenotemark{a}   &    H$\alpha$+[NII]: $236\pm3$  & H$\alpha$+[NII]: $36\pm9$  &  7.4     &
\vline \  A3638\tablenotemark{a}   &   H$\alpha$+[NII]: $171\pm3$  & H$\alpha$+[NII]: $-34\pm12$ & 8.6    \\
 A3806\tablenotemark{a}   &    H$\alpha$+[NII]: $85\pm5$  & H$\alpha$+[NII]: $-95\pm13$  &   5.9    &
\vline \  A3880\tablenotemark{a}   &   H$\alpha$+[NII]: $255\pm3$  & H$\alpha$+[NII]: $-35\pm11$ &  13.2   \\
 A3998\tablenotemark{a}   &   H$\alpha$+[NII]: $123\pm4$  & H$\alpha$+[NII]: $-30\pm11$  &   17.8    &
\vline \  A4059\tablenotemark{a}   &   H$\alpha$+[NII]: $203\pm3$  & H$\alpha$+[NII]: $-65\pm10$ &  7.9    \\
 A478\tablenotemark{a}     & H$\alpha$+[NII]: $129\pm4$  & H$\alpha$+[NII]: $-126\pm15$  &  17.0     &
\vline \  A496\tablenotemark{a}     &  H$\alpha$+[NII]: $125\pm4$  & H$\alpha$+[NII]: $0\pm12$  & 8.1   \\
 A85\tablenotemark{a}       &H$\alpha$+[NII]: $149\pm4$  & H$\alpha$+[NII]: $-172\pm12$  &   8.3    &
\vline \  Hydra-A\tablenotemark{a}  &     H$\alpha$+[NII]: $211\pm3$  & H$\alpha$+[NII]: $-102\pm10$ & 7.9    \\
 N4325\tablenotemark{a}  &     H$\alpha$+[NII]: $110\pm4$  & H$\alpha$+[NII]: $-40\pm13$  &  6.7     &
\vline \  Rc0120\tablenotemark{a}  &     H$\alpha$+[NII]: $182\pm3$  & H$\alpha$+[NII]: $-38\pm10$ &  4.7   \\
 Rc1524\tablenotemark{a} &     H$\alpha$+[NII]: $192\pm3$  & H$\alpha$+[NII]: $0\pm9$  &  15.6     &
\vline \  Rc1539\tablenotemark{a} &     H$\alpha$+[NII]: $187\pm3$  & H$\alpha$+[NII]: $80\pm14$ & 13.7    \\
 Rc1558\tablenotemark{a} &     H$\alpha$+[NII]: $138\pm3$  & H$\alpha$+[NII]: $-66\pm33$  &  13.8     &
\vline \  Rc2101\tablenotemark{a} &     H$\alpha$+[NII]: $88\pm5$  & H$\alpha$+[NII]: $0\pm24$ &  7.1   \\
 R0000\tablenotemark{a} &     H$\alpha$+[NII]: $144\pm4$  & H$\alpha$+[NII]: $-17\pm11$  &   3.5    &
\vline \  R0338\tablenotemark{a} &     H$\alpha$+[NII]: $190\pm3$  & H$\alpha$+[NII]: $15\pm10$ &  8.6   \\
 R0352\tablenotemark{a} &     H$\alpha$+[NII]: $190\pm3$  & H$\alpha$+[NII]: $-23\pm14$  &  17.8     &
\vline \  R0747\tablenotemark{a} &     H$\alpha$+[NII]: $191\pm3$  & H$\alpha$+[NII]: $39\pm4$ &  24.8   \\
 R0821\tablenotemark{a} &     H$\alpha$+[NII]: $122\pm4$  & H$\alpha$+[NII]: $20\pm13$  &   20.4    &
\vline \  S555\tablenotemark{a}     &     H$\alpha$+[NII]: $232\pm3$  & H$\alpha$+[NII]: $61\pm10$ & 11.5    \\
 Rc1436\tablenotemark{a}  &     H$\alpha$+[NII]: $122\pm4$  & H$\alpha$+[NII]: $68\pm11$  & 14.6      &
\vline \  A1111\tablenotemark{a}    &     H$\alpha$+[NII]: $113\pm3$  & H$\alpha$+[NII]: $-277\pm12$ & 21.9   \\
 A1204\tablenotemark{a}    &     H$\alpha$+[NII]: $221\pm3$  & H$\alpha$+[NII]: $109\pm15$  &  26.3     &
\vline \  A2390\tablenotemark{a}    &     H$\alpha$+[NII]: $231\pm2$  & H$\alpha$+[NII]: $-91\pm35$ &   26.0  \\
 A3378\tablenotemark{a}    &     H$\alpha$+[NII]: $80\pm4$  & H$\alpha$+[NII]: $-54\pm12$  &  20.6     &
\vline \  A3639\tablenotemark{a}    &     H$\alpha$+[NII]: $157\pm3$  & H$\alpha$+[NII]: $-56\pm40$ &  18.8   \\
 A383\tablenotemark{a}      &     H$\alpha$+[NII]: $244\pm2$  & H$\alpha$+[NII]: $-178\pm46$  &   24.2    &
\vline \  Rc0132\tablenotemark{a} &     H$\alpha$+[NII]: $205\pm3$  & H$\alpha$+[NII]: $-82\pm8$ &  24.8   \\
 Rc0331\tablenotemark{a} &     H$\alpha$+[NII]: $183\pm3$  & H$\alpha$+[NII]: $-128\pm13$  &   14.5    &
\vline \  Rc0944\tablenotemark{a} &     H$\alpha$+[NII]: $185\pm4$  & H$\alpha$+[NII]: $-304\pm11$ &  25.5   \\
 Rc2014\tablenotemark{a} &     H$\alpha$+[NII]: $206\pm3$  & H$\alpha$+[NII]: $53\pm32$  &   18.6    &
\vline \  Rc2129\tablenotemark{a} &     H$\alpha$+[NII]: $114\pm3$  & H$\alpha$+[NII]: $158\pm11$ & 30.5    \\
 R1651\tablenotemark{a} &     H$\alpha$+[NII]: $177\pm3$  & H$\alpha$+[NII]: $-12\pm10$  &   17.5    &
\vline \  S780\tablenotemark{a}     &     H$\alpha$+[NII]: $188\pm2$  & H$\alpha$+[NII]: $258\pm10$ &  33.0   \\
 Z3179\tablenotemark{a}   &     H$\alpha$+[NII]: $80\pm8$  & H$\alpha$+[NII]: $-66\pm17$  & 14.8      &
\vline \  A3444\tablenotemark{a} &  H$\alpha$+[NII]: $133\pm3$ & H$\alpha$+[NII]: $-80\pm16$ &  35.7  \\
  R0439\tablenotemark{a}   &     H$\alpha$+[NII]: $184\pm3$  & H$\alpha$+[NII]: $76\pm11$  &   26.1    & 
\vline \  A795\tablenotemark{a}     &     H$\alpha$+[NII]: $309\pm2$  & H$\alpha$+[NII]: $-146\pm7$ &  11.1  \\
    A3017\tablenotemark{a}   &     H$\alpha$+[NII]: $189\pm3$  & H$\alpha$+[NII]: $-333\pm9$  &  25.0     & 
\vline \  A1795\tablenotemark{b}  &  H$\alpha$: $205\pm1$ & H$\alpha$: $60\pm11$ &  50  \\   
 A1991\tablenotemark{a}   &     H$\alpha$+[NII]: $  109\pm4$  & H$\alpha$+[NII]: $-107\pm14$  &  16.4     &       
\vline \  N1275\tablenotemark{e} & H$\alpha$+[NII]: $137\pm20$     & H$\alpha$+[NII]:: $-43\pm32$ &  45  \\
 A2597\tablenotemark{b}  &  H$\alpha$: $241\pm1$ & H$\alpha$: $76\pm10$ &  17  &
\vline \ Se15903\tablenotemark{b}  &  H$\alpha$: $160\pm2$ & H$\alpha$: $-121\pm12$ &   23    \\      
  M87\tablenotemark{f} &  [CII]: $153\pm11$     &  [CII]: $-62\pm11$ &   3.8 &
\vline \   Rc1257\tablenotemark{a} &     H$\alpha$+[NII]: $128\pm2$  & H$\alpha$+[NII]: $0\pm3$  &   6.8  \\
 N5044\tablenotemark{a,d}  &  H$\alpha$+[NII]: $190\pm2$  & H$\alpha$+[NII]: $-77\pm4$  &  5.6   &                                            
\vline \  Rc1511\tablenotemark{a} &     H$\alpha$+[NII]: $208\pm2$  & H$\alpha$+[NII]: $-7\pm2$ & 5.0     \\                                             
                                            &  CO: $177\pm10$ & CO: $-169\pm8$ &  4   & 
\vline \  Rc1304\tablenotemark{a} &     H$\alpha$+[NII]: $176\pm2$  & H$\alpha$+[NII]: $-39\pm3$ &  3.3   \\
 A3574\tablenotemark{a}   &  H$\alpha$+[NII]: $106\pm2$  & H$\alpha$+[NII]: $-27\pm4$  &  1.5     &
\vline \  A194\tablenotemark{a}     &     H$\alpha$+[NII]: $90\pm1$  & H$\alpha$+[NII]: $19\pm3$  & 1.5   \\    
 S805\tablenotemark{a}     &     H$\alpha$+[NII]: $121\pm2$  & H$\alpha$+[NII]: $35\pm4$  &   2.7    &
\vline \  S851\tablenotemark{a}     &     H$\alpha$+[NII]: $209\pm4$  & H$\alpha$+[NII]: $-23\pm8$ &   3.6  \\
 N4636\tablenotemark{c} & [CII]: $153\pm3$   & [CII]: $22\pm3$ &   1.8    &
\vline \  N533\tablenotemark{a}    &     H$\alpha$+[NII]: $138\pm3$  & H$\alpha$+[NII]: $-37\pm5$ & 2.8    \\
                                                       & [OI]: $99\pm12$    & [OI]: $75\pm12$ &  1     & 
\vline \  N5846\tablenotemark{a,c}  & H$\alpha$+[NII]: $118\pm2$  & H$\alpha$+[NII]: $30\pm4$  &  1.3  \\
 H62\tablenotemark{a}  &     H$\alpha$+[NII]: $103\pm2$  & H$\alpha$+[NII]: $-27\pm4$  &  3.3     &
\vline \                                        & [CII]: $202\pm4$ & [CII]: $-25\pm4$ &  2.5      \\         
 N5813\tablenotemark{a,c}   & H$\alpha$+[NII]: $133\pm6$  & H$\alpha$+[NII]: $48\pm15$ &  1.5   &
\vline \  N6868\tablenotemark{c}  & [CII]: $216\pm3$ & [CII]: $125\pm3$ &  2.8  \\
                                               & [CII]: $178\pm4$  & [CII]: $96\pm4$ &  3.7  & 
\vline \                                                 & [OI]: $215\pm13$ & [OI]: $137\pm13$ &  1  \\ 
                                               & [OI]: $116\pm15$  & [OI]: $30\pm15$ &   1    &  
\vline \  N7049\tablenotemark{c}  & [CII]: $168\pm3$ & [CII]: $78\pm3$ &  3.2 
\enddata
\tablenotetext{}{\small {\it Notes.} 
The ensemble ${\rm FWHM}/2.355=\sigma_{v,{\rm los}}$ and velocity offset (from the systemic velocity) $\bar{v}_{\rm los}$ are derived from the single integrated spectrum extracted within the detectable emission region $R_{\rm ex}$ (major axis length; \S\ref{s:obs_ens}); for clusters/groups the median is $R_{\rm ex}\simeq 14/3$\;kpc. All objects are galaxy clusters, except for the last 15 groups.
The rest-frame wavelengths of the tabulated lines are H$\alpha$ 6562.8\;\AA, [NII] 6548/6583\;\AA, [CII] $157.7\;\mu$m, [OI] $63.2\;\mu$m, and CO(2-1) $1.3$\;mm.
Where the line shift is consistent with no offset from the systemic velocity, a null shift is reported. 
The object prefixes A, S, H, N, R/Rc, Se, and Z are abbreviations for the Abell, Abell Supplementary, HCG, NGC, RXJ/RXCJ, Sersic, and Zwicky catalogs, respectively.
References:
{\emph{a)}} Newly computed from \citet{Hamer:2016} VIMOS IFU (VLT) data. 
{\emph{b)}} Newly computed from \citet{McDonald:2012_Ha} {\it Magellan} and {\it Keck} data.
{\emph{c)}} {\it Herschel} data from \citet{Werner:2014}. 
{\emph{d)}} Newly computed from ALMA data (\citealt{Temi:2017}; velocity offsets dispersion method). 
{\emph{e)}} Newly computed from SITELLE (CFHT) data (Gendron-Marsolais et al.~in prep.); $R<4$\;kpc excised.
{\emph{f)}} {\it Herschel} data from \citet{Werner:2013}. 
\\}
\label{t:ens}
\end{deluxetable*}
\end{center}
\capstarttrue

\vspace{-0.8cm}
\section{Observational Data and Comparison with Simulations} \label{s:obs}
\noindent
We present here new observational constraints -- together with the available literature data --  on the line broadening and line shift of the warm and cold gas kinematics in massive galaxies mainly within the cores of clusters and groups (spanning the mass range $M_{500}\sim10^{13}$\,-\,$10^{15}\;\msun$). We compare them with the above numerical results, discussing the key differences between the ensemble and pencil-beam method, as well as the related limitations. 
At the same time, the following data points provide an estimate for turbulence velocities and local cloud kinematics, which can be used in subsequent analytical, numerical, or observational works. For instance, this can be used to calibrate and remove systematics of indirect methods estimating the gas kinematics, or to model non-thermal pressure support in semianalytic studies and subgrid models for large-scale simulations.
We remark that the goal is to show the potential for such a kind of global analysis and not to delve into the details of each object, which is left to future work.

\vspace{+0.15cm}
\subsection{Ensemble-beam detections} \label{s:obs_ens}
\noindent
As shown in \S\ref{s:G12}-\ref{s:bro}, the primary method which allows us to retrieve the volume-filling turbulent velocity is to analyze the ensemble LOS velocity dispersion. In observations, this can be achieved by extracting a single, integrated spectrum over the maximally feasible radial aperture covering the entire warm ($\sim$\,$10^3$\,-\,$10^5$\;K) or cold ($\lta 200$\;K) gas emission region. To show such capability, we applied the ensemble method to the \citeauthor{Hamer:2016} (\citeyear{Hamer:2016} -- H16) sample including 68 well-resolved H$\alpha$+[NII] objects -- mostly galaxy cluster cores, plus 12 massive groups. An analogous method is applied to other 8 objects not included in the H16 sample and displaying cold/warm gas emission (e.g., \citealt{McDonald:2012_Ha, Werner:2014}; and the Perseus cluster, Gendron-Marsolais et al.~in prep.). 
Table~\ref{t:ens} lists the retrieved constraints and sample details for all the 76 objects.

The observational analysis to retrieve the gas FWHM ($\sigma_{v,{\rm los}}$) and line offset from the systemic velocity (shift) was carried out as follows -- by taking the H16 sample as reference. 
Initially, the total spectra of the continuum and line-emitting regions are extracted from the data cube (e.g., VIMOS IFU for the H16 sample).  For the line emission, this is done by taking the H$\alpha$ flux maps (emission detected at S/N\,$>7$). A masked cube is then created by discarding all spaxels in each wavelength channel with no emission.  The remaining spaxels are summed to give a total flux value for that channel, producing a total spectrum for the line-emitting regions. Table~\ref{t:ens} lists the extraction radius\footnote{Although representative of the warm gas bulk emission with high S/N, in a few objects, $R_{\rm ex}$ may not match the full extent of the nebula due to the limited field of view and/or H$\alpha$ absorption.}, with a median $R_{\rm ex}\simeq 14$\;kpc for clusters and 3\;kpc for groups.

The total continuum spectrum is determined in the same way, using a collapsed white-light image in place of the H$\alpha$ map and discarding spaxels where the continuum flux is less than 1/10 of the peak flux from the galaxy center\footnote{This empirical threshold ensures that sufficient sky pixels are recovered to calculate the sky emission, which is then subtracted.
We tested different thresholds (1/5 and 1/20) and found no significant difference in the retrieved velocity offsets.}.
The H$\alpha$ masking is then inverted (discarding only spaxels with H$\alpha$ detection) and the standard deviation of the remaining spaxels in each channel is calculated to provide an estimate of the error related to the two total spectra.
The kinematics of the ionized gas and stellar components of the galaxy are then determined by fitting Gaussian profiles to the relevant total spectrum (a triplet fit to the H$\alpha$+[NII] complex for the line-emitting gas and a negative doublet fit to the NaD absorption\footnote{We note NaD can have both stellar and gaseous origins; however, the NaD features we fit are all substantially broader than the emission lines, indicating they do not originate from the ionized gas in the galaxy and thus are most likely of stellar origin.} 
for the stellar component) using a $\chi^2$ minimization procedure (see H16 for more details).
Finally, the FWHM of the line-emitting gas is extracted directly from the fit\footnote{
We tested an alternative approach, computing the ensemble velocity dispersion as the RMS of the projected velocity shifts from several patches within $R_{\rm ex}$. This method introduces substantial noise due to patches with low signal. Moreover, each velocity shift has experienced positive/negative summation along LOS, so this RMS is typically a lower limit to the actual line broadening $\sigma_{v,{\rm los}}$. We recommend to use the more robust single-spectrum FWHM.
}.
The velocity difference between the H$\alpha$ emission and the NaD absorption gives the velocity offset. 
It is worth noting that the total uncertainty is dominated by the systematic error (spectral sampling and instrumental
line spread function), while random noise is drastically reduced due to the aggregation of several 100 spectra for each source (${\rm S/N}\propto N/\sqrt{N}$, with $N$ the number of spectra).

\begin{figure}
\subfigure{\includegraphics[width=1.005\columnwidth]{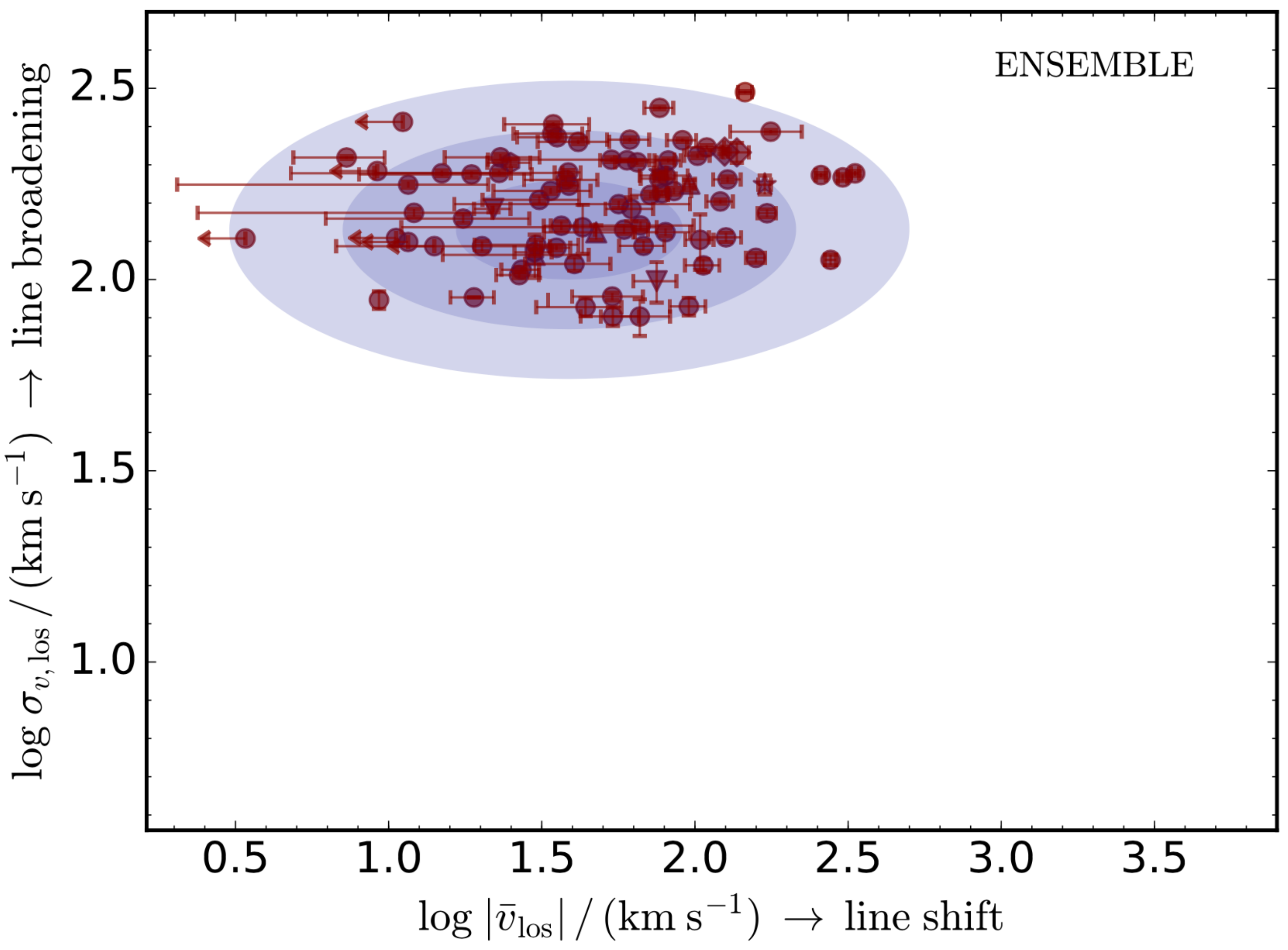}}
\subfigure{\includegraphics[width=1.005\columnwidth]{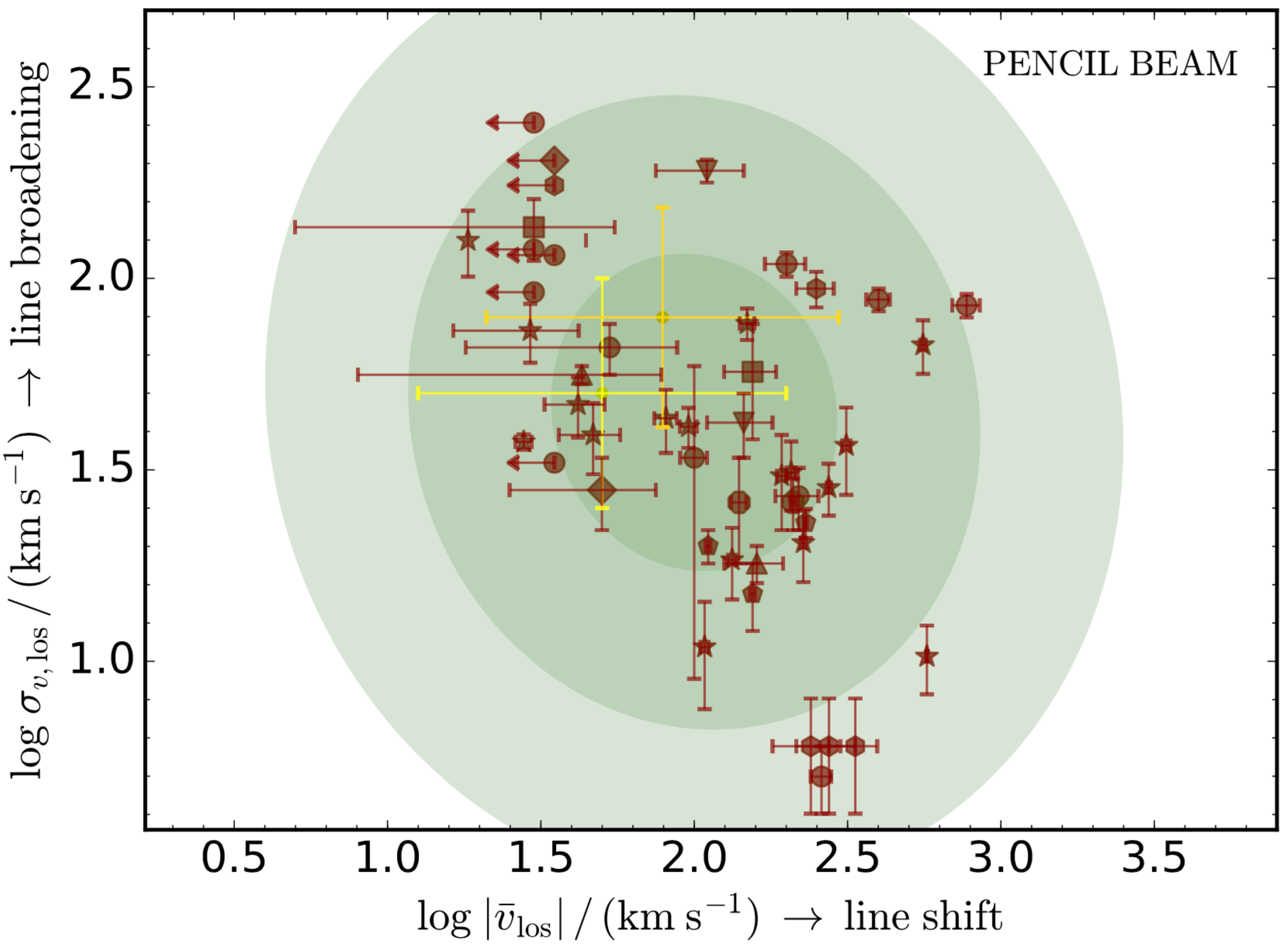}}
\caption{LOS velocity dispersion (line broadening) versus the magnitude of the LOS velocity (line shift) in logarithmic space for the warm and cold gas phases: comparison between the observational data (red points; Tab.~\ref{t:ens}-\ref{t:pen}) and the predictions from the simulations (contours) for the ensemble (top) and pencil-beam detections (bottom). The blue/green contours show the 1--3$\sigma$ confidence intervals (via covariance analysis) tied to the global lognormal distributions found in the simulations (\S\ref{s:G12}-\ref{s:G17}). Red arrows mark the data points that have velocity offset comparable to errors; multiple detections in the same object are marked with identical non-circle symbols. The yellow and orange bars mark the logarithmic mean and deviation of the \citet{Maccagni:2017} sample and the Perseus regions (\citealt{Salome:2006,Salome:2008}), respectively.
Observations and data are consistent: the ensemble (several kpc aperture) measurement substantially reduces the scatter in the line broadening (with relatively small shifts), making the hot gas turbulence estimate reliable. The pencil-beam detections show substantial scatter, with typically lower line broadening and larger line shift; the wide range in velocity shifts allows dual components in the energy spectrum to be detected.
\\
} 
\label{f:comp}
\end{figure}

As a diagnostic tool to understand the kinematic properties with different methods, we propose a novel diagram confronting the line broadening versus the line shift magnitude. Fig.~\ref{f:comp} shows the observed data points for warm and cold gas, which are compared with the simulation predictions (shaded contours). The shaded contours show the simulation bivariate distributions with mean and 1-3 standard deviations found in the previous run (\S\ref{s:G12}),
which are best fitted by lognormal distributions.
As the \S\ref{s:G12} run probes a large dynamical range and varying cluster regimes,
we use as reference line-broadening normalization the mean of its entire hot gas distribution, $\sigma_{v,{\rm los, hot}}\simeq140\;\kms$ (which is comparable to that of the warm gas; Fig.~\ref{f:G12_2corr}), and show the logarithmic standard deviation from such a mean. The simulated global velocity offsets are discussed in \S\ref{s:shift}. 

The top panel in Fig.~\ref{f:comp} shows that the ensemble detection -- both in simulations and observations -- cover a specific section of the $\log \sigma_{v,{\rm los}} - \log |\bar{v}_{\rm los}|$ diagram, namely the top-left region, which is the locus of substantial line broadening and relatively low velocity offsets. The scatter in velocity dispersion is mild due to the ensemble/single-spectrum approach, which decreases the statistical noise of single clouds and filaments. 
The log mean of the two distributions differs by 3\% and 6\% along the broadening and shift axis, respectively. The overlap within 2$\sigma$ is evident to the naked eye, with essentially a null correlation angle and analogous symmetry.
Along the broadening/shift direction, the observed data have a mildly larger standard deviation (5\%/6\%). 
Given the limitations of the models and observational biases, we deem the simulations and observations to be in good agreement. 
E.g., the observed velocity offsets can be very sensitive to the redshift measurement of the host galaxy, with systematics not always captured.
Further, coherent warm gas structures dominating the field of view have the tendency to increase the LW velocity offsets.
Running a 2D Kolmogorov-Smirnov (KS$_{\rm 2d}$)\footnote{
Following \citet{NRec:1992} based on \citet{Fasano:1987}.
The synthetic data sample is randomly extracted from the simulation distribution with the same number of data points as that of the observational sample (excluding the few upper limits) and bootstrapped 1000 times to give a mean $p$-value. As with any single likelihood value, this should be taken with a grain of salt. First, the KS statistic in 2D is only approximated (as cumulative distribution functions are not well-defined in more than 1D). Second, the number of current observed points is limited. Perhaps more important, a single value attached to a comparison 
is reductive of the complexity included in either the simulation or observational measurement.} 
two-sample test results in a $p$-value of 0.02, implying that the null hypothesis cannot be rejected at the customary confidence levels of 99\% and 99.9\%.
This corroborates the visual inspection that the two datasets do not deviate by a large amount
but are perhaps not drawn from an identical distribution. 
Nevertheless, perfect equivalence shall not be expected considering the detailed differences between the synthetic and real value measurements.

The above analysis benefiting from a large (76) sample size shows that turbulence in galaxy cluster/group cores is contained within a relatively narrow window, $\sigma_{v,{\rm los}}\approx 100$\,-\,250$\;\kms$, which implies subsonic Mach numbers ${\rm Ma_{\rm 1d}}\sim0.1$\,-\,0.3. This corroborates indirect X-ray hot gas estimates (\S\ref{s:intro}); e.g., \citet{Hofmann:2016} and \citet{Zhuravleva:2017} find a similar range via plasma density fluctuations in the core, with $\sigma_{v,{\rm los}}$ within 50\;$\kms$ from our values (interesting cases are A2052, A85, and A1795), although their uncertainties remain substantial because of the density contamination tied to substructures. The upper and lower limits set via XMM-RGS (e.g., \citealt{Ogorzalek:2017}) and cosmological simulations (e.g., \citealt{Lau:2009,Lau:2017,Nagai:2013}) further support such range of subsonic ICM/IGrM turbulence.
Last but not least, it is remarkable that the SITELLE warm gas constraint matches the {\it direct} high-resolution $\sigma_{v, {\rm los}}$ observed by {\it Hitomi} in the archetypal galaxy cluster Perseus (\S\ref{s:G12}).
While waiting for the next-generation X-ray instruments (e.g., XARM -- the successor of {\it Hitomi} -- and {\it Athena}; \citealt{Ettori:2013})
to give us high-precision turbulence detections, this analysis allows us to use
warm and cold gas velocity dispersions as robust proxies for the hot gas turbulent velocities in large samples (note that future X-ray instruments will still require several days of exposure for just one object).

For five objects (Tab.~\ref{t:ens}) we report literature detections in more than one band (besides Perseus, which was tackled in \S\ref{s:G12}). 
Except for NGC 5846, the four other galaxies have ensemble broadening comparable between the cold and warm gas within $\sim\,$50\;$\kms$, with NGC 5044 and NGC 6868 two exemplary cases. The velocity offsets are also broadly aligned.
On the other hand, the currently available extraction regions are often dissimilar; indeed, we observe larger values, all associated with larger extraction regions. 
Upcoming systematic multiwavelength investigations -- which our team is currently undertaking with {\it Chandra}, XMM, ALMA, VLT, SOFIA, HST, {\it Magellan}, SOAR, SITELLE, and IRAM -- will be key to calibrate the multiphase kinematics over the same extraction region and with similar depth.
At the present, the ensemble H$\alpha+$[NII] nebulae appear to be the best channel to test the volume-filling turbulence and related velocity dispersion.

\capstartfalse
\begin{center}
\begin{deluxetable}{lll}
\vspace{-0.6cm}
\setlength{\tabcolsep}{2pt}
\tabletypesize{\small}
\tablecaption{Detected line broadening and shift for the {\it pencil-beam} cold/warm phase in observed massive galaxies within cluster/group cores. These constraints can be used in other studies requiring the kinematics of the small-scale filaments or accreting clouds.}
\setstretch{0.7}
\tablehead{
\colhead{Object \quad\quad\quad\quad} & \colhead{$\sigma_{v,{\rm los}}$ [$\kms$] \quad\quad} & \colhead{\quad$\bar{v}_{\rm los}$ [$\kms$]\quad\quad}
}
\tablecolumns{3}
\startdata
A2390 absor.~broad\tablenotemark{a} & HI: $191\pm13$ & HI: $-110\pm35$  \\
\ \ \ \ \ \ \ \ \;\,absor.~narrow\tablenotemark{a} & HI: $42\pm8$ & HI: $145\pm35$  \\
Z8276 absorption\tablenotemark{a}             & HI: $27\pm5$ & HI: $219\pm35$  \\
A1795 absorption\tablenotemark{a}            & HI: $255\pm21$ & HI: $0\pm30$  \\
Z8193 absorption\tablenotemark{a}            & HI: $92\pm8$ & HI: $0\pm30$  \\
N6338 absorption\tablenotemark{a}     & HI: $109\pm8$ & HI: $-200\pm30$  \\
R1832 absorption\tablenotemark{a}       & HI: $85\pm6$ & HI: $-773\pm80$  \\
R1558 absor.~broad\tablenotemark{a}  & HI: $56\pm3$ & HI: $43\pm35$  \\
\ \ \ \ \ \ \ \ \;\,absor.~narrow\tablenotemark{a} & HI: $18\pm2$ & HI: $160\pm35$  \\
R1350 absorption\tablenotemark{a}       & HI: $119\pm8$ & HI: $0\pm30$  \\
A2597 absor.~broad\tablenotemark{a}         & HI: $175\pm15$ & HI: $0\pm35$  \\
\ \ \ \ \ \ \ \ \;\,absor.~narrow\tablenotemark{a}        & HI: $94\pm10$ & HI: $250\pm35$  \\
\ \ \ \ \ \ \ \ \;\,absor.~narrow\tablenotemark{b} & CO: $6\pm 2$ & CO: $240\pm 60$    \\
\ \ \ \ \ \ \ \ \;\,absor.~narrow\tablenotemark{b} & CO: $6\pm 2$ & CO: $275\pm 60$   \\
\ \ \ \ \ \ \ \ \;\,absor.~narrow\tablenotemark{b}\quad\quad & CO: $6\pm 2$ & CO: $335\pm 60$   \\
N1275 absor.~broad\tablenotemark{a}    & HI: $203\pm15$ & HI: $0\pm35$  \\
\ \ \ \ \ \ \ \ \;\,absor.~narrow\tablenotemark{a}  & HI: $28\pm6$ & HI: $50\pm25$  \\
PKS1353 absorption\tablenotemark{a}            & HI: $66\pm10$ & HI: $-53\pm35$  \\
Cygnus-A absorption\tablenotemark{a}            & HI: $115\pm8$ & HI: $0\pm35$  \\
Hydra-A absorption\tablenotemark{a}            & HI: $33\pm4$ & HI: $0\pm35$  \\
4C55.16 absorption\tablenotemark{c}            & HI: $88\pm6$ & HI: $-399\pm35$  \\ 
R1603 absor.~broad\tablenotemark{a}    & HI: $136\pm25$ & HI: $-30\pm25$  \\
\ \ \ \ \ \ \ \ \;\,absor.~narrow\tablenotemark{a}  & HI: $57\pm19$ & HI: $-155\pm30$  \\
N4636 emission\tablenotemark{d} & CO: $26\pm8$ & CO: $140\pm8$  \\
\ \ \ \ \ \ \ \ \;\,emission\tablenotemark{d} & CO: $26\pm4$ & CO: $210\pm4$  \\
N5846 emission\tablenotemark{d}             & CO: $23\pm2$ & CO: $-231\pm2$  \\
\ \ \ \ \ \ \ \ \;\,emission\tablenotemark{d}& CO: $15\pm3$ & CO: $-155\pm3$  \\
\ \ \ \ \ \ \ \ \;\,emission\tablenotemark{d} & CO: $20\pm2$ & CO: $111\pm2$  \\
N5044 absorption\tablenotemark{e}            & CO: $5\pm1$ & CO: $260\pm20$  \\
\ \ \ \ \ \ \ \ \;\,emission\tablenotemark{d}   & CO: $126 \pm 25 $ & CO: $0 \pm 26 $ \\
\ \ \ \ \ \ \ \ \;\,emission\tablenotemark{d}   & CO: $76 \pm 7 $ & CO: $-149 \pm 8 $ \\
\ \ \ \ \ \ \ \ \;\,emission\tablenotemark{d}   & CO: $73 \pm 13 $ & CO: $29 \pm 13 $ \\
\ \ \ \ \ \ \ \ \;\,emission\tablenotemark{d}   & CO: $67 \pm 11$ & CO: $-557 \pm 12 $ \\
\ \ \ \ \ \ \ \ \;\,emission\tablenotemark{d}   & CO: $47 \pm 8 $ & CO: $42 \pm 9 $ \\
\ \ \ \ \ \ \ \ \;\,emission\tablenotemark{d}   & CO: $43 \pm 8 $ & CO: $-81 \pm 7 $ \\
\ \ \ \ \ \ \ \ \;\,emission\tablenotemark{d}   & CO: $41 \pm 5 $ & CO: $-96 \pm 6 $ \\
\ \ \ \ \ \ \ \ \;\,emission\tablenotemark{d}   & CO: $39 \pm 8 $ & CO: $-47 \pm 11 $ \\
\ \ \ \ \ \ \ \ \;\,emission\tablenotemark{d}   & CO: $37 \pm 2 $ & CO: $28 \pm 2 $ \\
\ \ \ \ \ \ \ \ \;\,emission\tablenotemark{d}   & CO: $37 \pm 9 $ & CO: $-313 \pm 9 $ \\
\ \ \ \ \ \ \ \ \;\,emission\tablenotemark{d}   & CO: $31 \pm 6 $ & CO: $-207 \pm 7 $ \\
\ \ \ \ \ \ \ \ \;\,emission\tablenotemark{d}   & CO: $30 \pm 8 $ & CO: $-193 \pm 7 $ \\
\ \ \ \ \ \ \ \ \;\,emission\tablenotemark{d}   & CO: $28 \pm 4 $ & CO: $-274 \pm 4 $ \\
\ \ \ \ \ \ \ \ \;\,emission\tablenotemark{d}   & CO: $20 \pm 4 $ & CO: $-227 \pm 4 $ \\
\ \ \ \ \ \ \ \ \;\,emission\tablenotemark{d}   & CO: $18 \pm 4 $ & CO: $-133 \pm 4 $ \\
\ \ \ \ \ \ \ \ \;\,emission\tablenotemark{d}   & CO: $11 \pm 3 $ & CO: $-108 \pm 3 $ \\
\ \ \ \ \ \ \ \ \;\,emission\tablenotemark{d}   & CO: $10 \pm 2 $ & CO: $-574 \pm 3 $ \\
A3716 absorption\tablenotemark{f}    & NaD: $34\pm25$ & NaD: $100\pm10$  \\
Perseus emission\tablenotemark{g} [log] & $\langle$CO$\rangle$: $1.9\pm0.3$  &  $\langle$CO$\rangle$: $1.9\pm0.4$  \\
66 radio galaxies\tablenotemark{h} [log] & $\langle$HI$\rangle$: $1.7\pm0.3$  &  $\langle$HI$\rangle$: $1.7\pm0.6$   
\enddata
\tablenotetext{}{\small {\it Notes.} 
Analogue of Table~\ref{t:ens}.
The rest-frame wavelengths  are HI\;$21$\,cm, CO(2-1)\;$1.3$\,mm, and NaD\;5890/5896\,\AA. 
Where the shift is consistent with no offset from the systemic velocity, null shift is reported. 
Except for A3716, all absorptions are against the radio AGN.
References:
{\emph{a)}} VLA, WRST, and ATCA data from \citeauthor{Hogan:2014} (\citeyear{Hogan:2014} and refs.~within). 
{\emph{b)}} ALMA data from \citet{Tremblay:2016}.
{\emph{c)}} WRST data from \citet{Vermeulen:2003}.
{\emph{d)}} Newly computed from ALMA CO(2-1) center and off-center emission (\citealt{Temi:2017}): N4636 and N5846 are new ALMA Cycle 3 observations, while N5044 is newly reduced from Cycle 0 data (with S/N $\ge6$).
{\emph{e)}} ALMA data from \citet{David:2014}. 
{\emph{f)}} MUSE data from \citet{Smith:2017}: NaD absorption against the stellar light of A3716 BCG (central E sector).
{\emph{g)}} IRAM low-resolution data (159 regions; log mean and RMS) from \citet{Salome:2006,Salome:2008}.
{\emph{h)}} Log mean and RMS of HI absorbers for WRST \citet{Maccagni:2017} sample (non-central radio galaxies).
\\}
\label{t:pen}
\end{deluxetable}
\end{center}
\capstarttrue

\vspace{-0.4cm}
\subsection{Pencil-beam detections} \label{s:obs_pen}
\noindent
For the pencil-beam (small aperture, below a few arcsec) detections in massive galaxies, we report the published value from the literature and a few new detections.
A larger sample requires new observational programs (e.g., one recently approved in ALMA Cycle 5 -- PI: A.~Edge).
The most used pencil-beam approach is to detect absorption lines (e.g., due to HI and CO clouds having significant optical depth) against the (radio) AGN, which acts as a backlight source. 
The BCG stellar light can also be used as backlight, in combination with NaD absorption.
The analysis procedure is then analogous to that above, extracting the systemic velocity offset and FWHM from Gaussian fitting of the absorption lines in the continuum-subtracted spectrum. \citet{Hogan:2014} discusses in detail the observational reduction with different instruments such as VLA and WRST (see \citealt{Tremblay:2016} for ALMA data). Although more challenging due to the low S/N in massive galaxies, small-aperture observations can be used to track small-scale clouds via emission features such as CO(2-1).
Table~\ref{t:pen} lists the 47 fiducial detections for the 19 available objects, with the addition of the mean properties of the \citeauthor{Maccagni:2017} (\citeyear{Maccagni:2017} -- M17) radio galaxy sample and of the Perseus IRAM regions. 

Fig.~\ref{f:comp} (bottom panel) shows the observational detections as red points (identical non-circle symbols denote the same host galaxy), compared with the simulation results for all the condensed gas with 1--3$\sigma$ confidence intervals. For the pencil-beam approach, the \S\ref{s:G17} run covers a meaningful range to statistically test the broadening distribution; the reference mean $\sigma_{v,{\rm los,hot}}$ is the same as in the previous section, but now the condensed gas has a ratio lower than 1:1 due to the pencil-beam sampling (Fig.~\ref{f:G17_broad}, bottom). The simulated velocity offsets over all of the condensed gas are discussed in \S\ref{s:shift}.

As anticipated by the numerical analysis, it is clear that this method substantially boosts the logarithmic scatter in both the broadening and shift dimensions up to nearly 0.5\;dex. The mean line broadening ($\simeq40\;\kms$) is significantly lower than the ensemble measurement, as predicted by the simulations. The mean velocity offset has instead larger values $\gta100$\;$\kms$. This is because the pencil beam is sampling smaller and typically fewer condensed elements. About 1/3\footnote{Multicomponent systems should be much easier to observe with the more modern receivers (e.g., CABB on ATCA), which combine a wide bandwidth with high resolution.} of the central galaxies observed in absorption display a dual component: either with a large line broadening and small shift or with a large shift and fairly contained broadening.
A mild anticorrelation appears to be present, with the faster (and denser) clouds associated with the nuclear inflow toward the SMBH sink region. 
It is interesting to observe that the broad component in absorption is often a reasonable proxy for the ensemble velocity dispersion, as it tends to sample multiple clouds along the LOS.
For instance, the pencil-beam broad components of A2390, A1795, A2597, and N1275 all reside within $\lta50\;\kms$ from the actual ensemble value, although there are exceptions, such as Hydra-A (which has an abnormal 5\;kpc disk).

In Fig.~\ref{f:comp} (bottom), we also plot the mean and deviation of the HI absorption detections in the M17 sample of 66 radio galaxies (yellow bars). Although comprising mostly non-central radio/elliptical galaxies and requiring deeper follow-up observations for each target, the HI absorption broadening and shift are consistent with the above central galaxies and simulation properties, with a slightly lower average line shift. An interesting case study is the early-type galaxy PKS 1718, where Maccagni et al. (in prep.) find redshifted clouds in HI, H$_2$, and CO clouds infalling within the Bondi radius with velocities up to $345\pm20\;\kms$, as found very similarly in A2597 central galaxy (\citealt{Tremblay:2016}).
Other interesting cases are PKS 1740 (\citealt{Allison:2015}), PKS 1657 (\citealt{Moss:2017}), and NGC 3998 (\citealt{Devereux:2018}).
This suggests that the top-down condensation and CCA are also likely common phenomena in low-mass/non-central galaxies, given that plasma atmospheres are expected to be ubiquitous (e.g., \citealt{Anderson:2015}) and subsonically perturbed by any AGN type (radio, quasar, etc.) and/or cosmic flow.

The pencil-beam method can be further used in emission and off center. For instance, taking advantage of ALMA high angular resolution and CO sensitivity, we probed several giant molecular associations in the massive galaxies NGC 4636, NGC 5846, and NGC 5044 (\citealt{Temi:2017}). Remarkably, the emission features show a similar mean and scatter to the above HI absorption features.
Commonly, small-scale clouds with low broadening ($< 35\;\kms$) are associated with large velocity offsets above 100\;$\kms$.
At the same time, the inspected masses, radii, and cospatiality of the giant molecular associations are consistent with the G17 simulation, corroborating the incidence of in-situ cooling via the multiphase cascade. 
Line absorption against the stellar light of a background galaxy is another promising way to retrieve the kinematics of gas at varying clustercentric distances (e.g., \citealt{Smith:2017}). 
Notice the beam must be small to achieve a proper pencil-beam detection (below the kpc or a few arcsec scale), otherwise the measurement will tend toward the ensemble approach; the IRAM data of Perseus (orange) is an example of intermediate regime with a beam of 4 kpc (12 arcsec).
 
Focusing on the comparison between the observed and simulated distributions in the $\log \sigma_{v,{\rm los}} - \log |\bar{v}_{\rm los}|$ diagram,
the mean differs by 2\% along both axes. The fairly good overlap within 2$\sigma$ is again evident to the naked eye.
The observed data have mildly larger/lower standard deviation (4\%/10\%) along the broadening/shift direction, respectively. 
Running a KS$_{\rm 2d}$ two-sample test (keeping in mind the limitations discussed in \S\ref{s:obs_ens}) results in a $p$-value of 0.14 (0.18 including the M17 and IRAM points). The absence of low $p$-values implies that the null hypothesis cannot be rejected, even at the less significant 95\% level. 
The point to take away is that the two distributions are similar, though not necessarily identical.
In particular, the simulation shows a milder anticorrelation (0.54\;rad difference in angle rotation), although
the scarcity of observed points in the bottom left section of the diagram may be attributed to the difficulty in detecting both small shift and narrow lines in the spectra.

Overall, granted that the sample requires larger statistics, both simulations and data agree well on the same picture:
the pencil-beam method is useful to track the inner infalling clouds (selecting the narrow features with a large velocity shift) or to have a preliminary estimate of the large-scale chaotic motions (selecting the broad features). The wide range of velocity shifts is key to allowing a clean separation of such components in the energy spectrum (e.g., in the radio band).
The ensemble measurement instead provides a robust and direct constraint on the volume-filling turbulent motions, as the ensemble condensed elements actively participate in the large-scale kinematics via the top-down multiphase condensation cascade.
The combination of the two approaches provides a powerful complementary diagnostic of the global and local gas kinematics.
\\
\\


\begin{figure*}
\centering
\subfigure{\includegraphics[width=0.98\columnwidth]{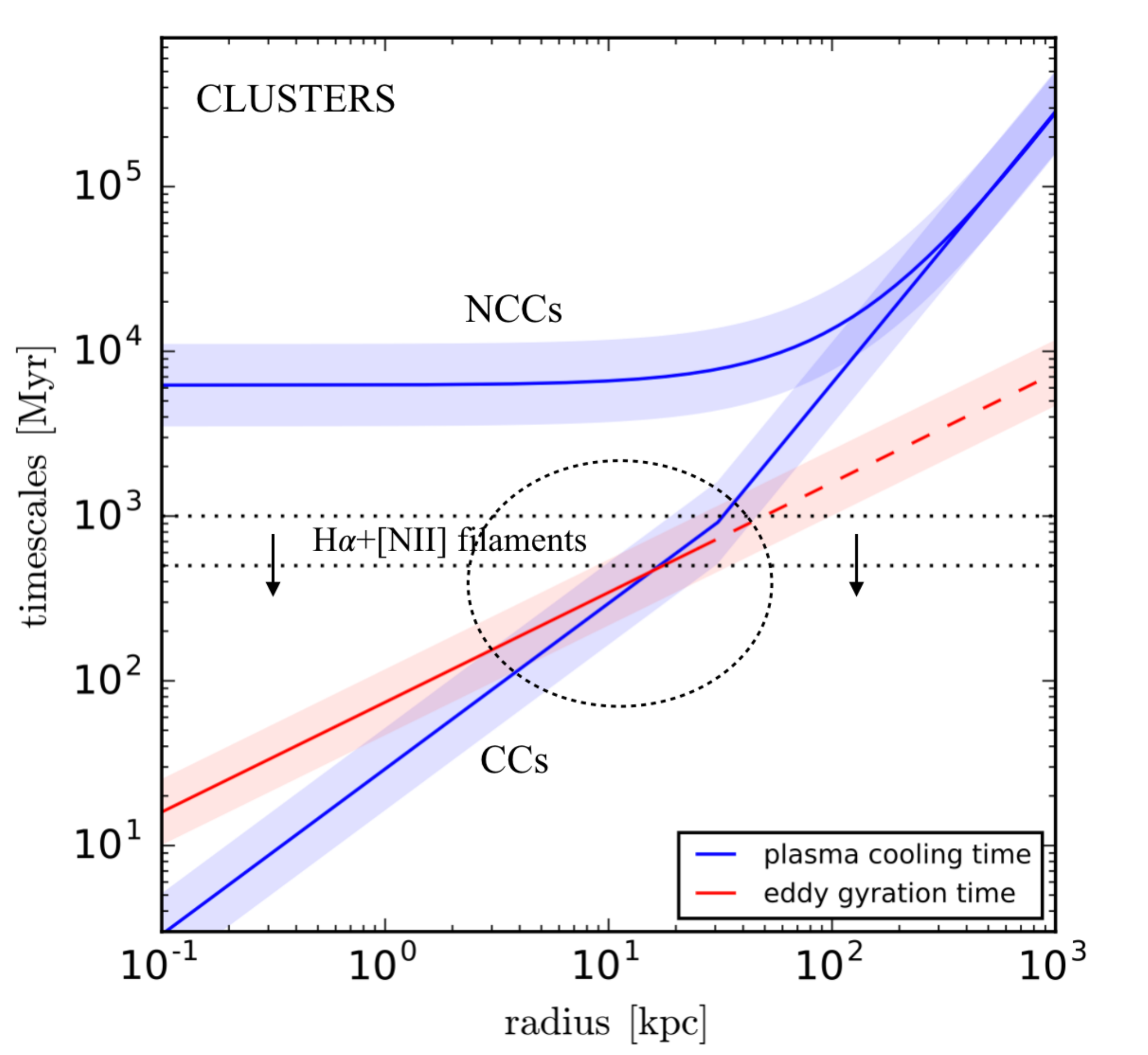}}
\subfigure{\includegraphics[width=0.98\columnwidth]{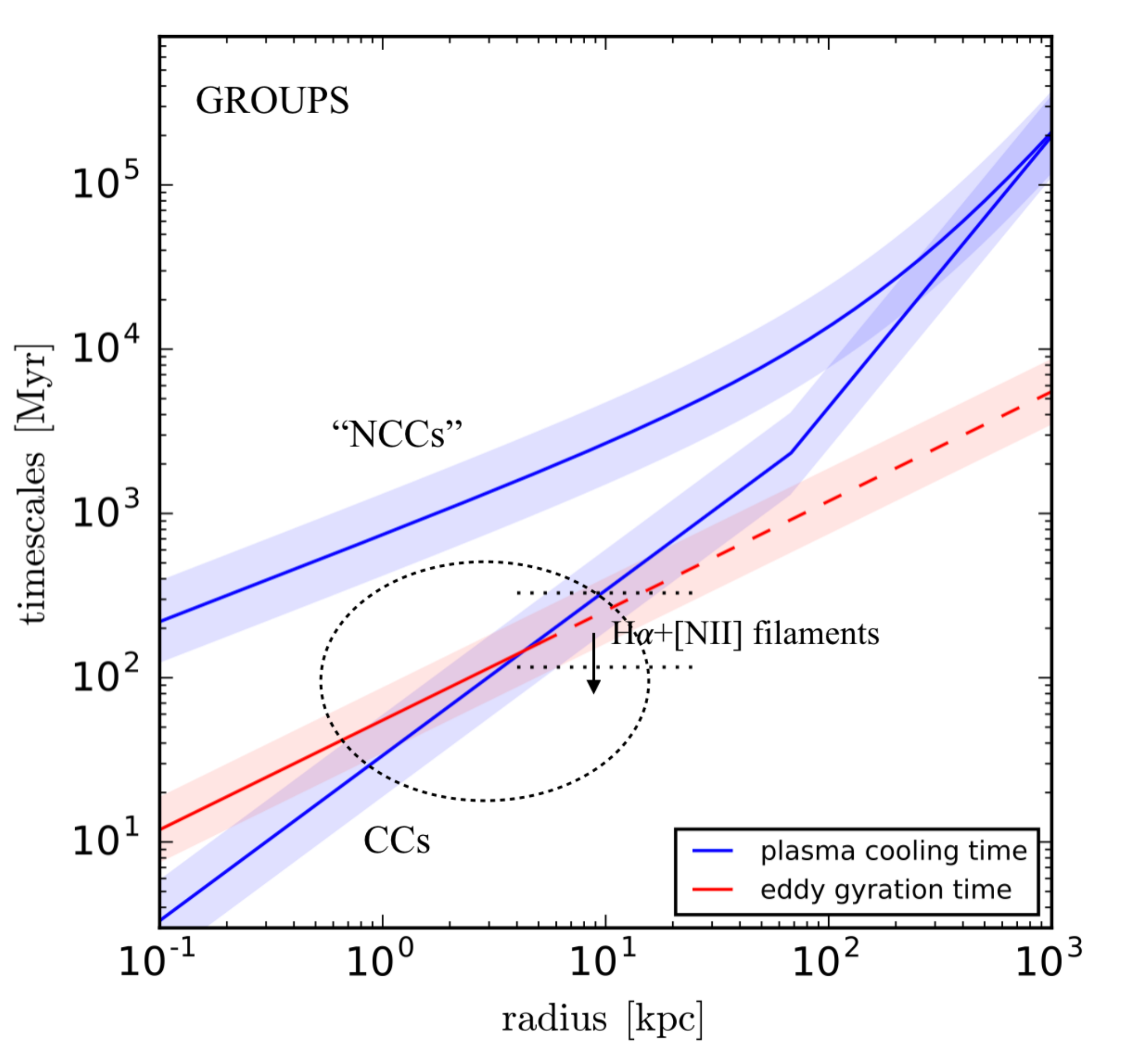}}
\caption{Dominant timescales related to the nonlinear multiphase condensation cascade ({\it left}: 5 keV cluster; {\it right}: 1 keV group).
{\it Left:} The blue curves are the mean plasma cooling times ($t_{\rm cool}\propto K^{3/2}$) of the observed non-cool-core (\citealt{Cavagnolo:2008}) and cool-core clusters (\citealt{Panagoulia:2014,Hogan:2017}). The red curve is the turbulence eddy turnover timescale ($t_{\rm eddy}\propto L^{1/3}\,\sigma_{v,L}^{-1}$; Eq.~\ref{e:teddy}).
The blue/red shaded bands mark the observed/simulated $\approx$\;90\% confidence dispersion. 
The horizontal dotted band separates the observed clusters with (below) or without (above) extended warm filaments (\citealt{Hogan:2017}).
The width of the dashed-ellipse indicates the range of extent radii found in filamentary warm gas detections (\citealt{McDonald:2010, McDonald:2011a}, and similarly in the simulations, \S\ref{s:obs_ens}); its height is related to the mean plasma $t_{\rm cool}$ at the extrema of such a range.
{\it Right}: Same as above but for 1 keV groups/massive galaxies. NCC groups with a flat entropy core seem nonexistent, so we plot as `NCCs' the upper envelope of observed groups from the most complete X-ray sample (\citealt{Sun:2009a}). At large radii, all profiles follow the adiabatic baseline and within $R_{500}$ the CC groups follow the \citet{Panagoulia:2014} scaling (\citealt{Werner:2014,Voit:2015_gE}).
Given the non-flattening entropy profiles of groups, the warm filament threshold is best taken within $r\simeq5$\,-\,15\;kpc. 
The eddy time (red) follows the AGN feedback constraints for groups (\citealt{Shin:2016}). 
The ratio $C\equiv t_{\rm cool}/t_{\rm eddy} \approx 1$ (in the range 0.6\,-\,1.8) marks well the condensation region over which clusters and groups display extended warm gas.
This criterion can be used in theoretical and observational works by leveraging the ensemble $\sigma_{v,{\rm los}}$ conversion between the hot, warm, and cold phase (\S\ref{s:res}).\\
\\
} 
\label{f:crit}
\end{figure*}

\section{The nonlinear multiphase condensation criterion} \label{s:crit}
\noindent
In the previous sections, we showed how the kinematics of the different phases is tightly related during the top-down condensation cascade and how we can convert between the ensemble velocity dispersions, in particular to that associated with the hot plasma. 
Here we discuss a key application of the retrieved turbulent velocities aimed to further understand the condensation process and the above multiwavelength observations.\\

\subsection{Previous condensation criteria}\label{s:pcrit}
 
A majorly debated topic in recent literature concerns the dominant criterion governing the formation of the condensed structures in the ICM/IGrM. Previous studies proposed that the ratio of the plasma cooling time to {\it free-fall} time\footnote{Defined as $t_{\rm ff}\equiv[2 r/g(r)]^{1/2}$, where $g(r)$ is the gravitational acceleration due to the total mass within a radius $r$.} falling below a few tens is the triggering threshold of thermal instability (a.k.a.~TI ratio). The mean value of the TI ratio is highly uncertain.
\citet{McCourt:2012} and \citet{Sharma:2012} propose a value of 10. \citet{Gaspari:2012a} simulations show a TI ratio between 8 and 25 (see their Fig.~3, bottom-right panel), which is similarly adopted by \citet{Voit:2015_nat}.
\citet{Valentini:2015} find TI-ratio thresholds that can reach a value of 70 in early-type galaxies.
Taken at face value and over different studies, the uncertainty on the TI ratio can be an order of magnitude.
It is important to note that this theoretical uncertainty accumulates on top of the intrinsic scatter due to system-to-system variations mainly tied to the cooling time (more in \S\ref{s:Ccrit}).

Four key issues arise with the TI ratio. 
First, the unity problem: why is the condensation occurring at such an elevated threshold, well above 1, which should be instead the physically sound transition\footnote{For example, in a steady spherical cooling flow, the Lagrangian entropy equation can be written as ${\rm d}\ln{K}/{\rm d}\ln{r} = t_{\rm in}/t_{\rm cool}$ (where $t_{\rm in}=r/v$), i.e., as the r.h.s.~ratio crosses below 1, condensation ensues (similar relations apply for the other conservation equations).}? 
Second, the large 1 dex theoretical uncertainty in the actual threshold -- even for similar systems -- hints at the fact that the free-fall time is not the primary timescale of the condensation process.
Third, observed condensed clouds do not follow ballistic orbits but are drifting with subvirial velocities (e.g., \citealt{McNamara:2014,Russell:2016}). 
Last but not least, there is the major observational hurdle of retrieving the free-fall times, i.e., the total masses, including the galactic and cluster baryonic and dark matter masses, which are notoriously challenging to constrain. 

Another suggested criterion (\citealt{Hogan:2017}) is to consider the cooling time below a fixed threshold, e.g., 1 Gyr. Although empirically effective and less noisy, this has the limitation of being applicable only to some classes of objects, e.g., massive clusters, but not groups or galaxies (e.g., \citealt{OSullivan:2017}). 
Finally, the uplift action of the AGN cavity may significantly increase the gas total inflow time ($t_{\rm in}$), well above the free-fall timescale, and bring it to altitudes where $t_{\rm cool}< t_{\rm in}$ (e.g., \citealt{McNamara:2016}).
However, it appears too restrictive to consider just one specific directional (bipolar) and short stage of the feedback cycle, namely the trailing phase.  The full process includes bubble inflation, uplift, and cocoon shocks, with isotropic turbulence being generated as a common by-product (this is true even if the driver is a merger or sloshing event). 
The turbulence time described below can be considered as a more universal concept of the reduced inflow time (with a clear quantitative observable), i.e., gas drifts in the halo regardless of the radial direction (inflow or outflow) and location (behind or around cavities).\\

\subsection{The new $C$-ratio criterion}\label{s:Ccrit}

Given the results in \S\ref{s:res}, namely the cold and warm gas following the progenitor chaotic kinematics, we show in Fig.~\ref{f:crit} that a robust and physically motivated condensation criterion is the ratio of the cooling time to the turbulence eddy turnover time $t_{\rm cool}/t_{\rm eddy} \approx 1$, which marks the multiphase state of the cluster/group core.
The condensation criterion can be alternatively written and interpreted as the velocity ratio $C=\sigma_v/v_{\rm cool}$, similar to a new dimensionless `Mach' number, where $v_{\rm cool}\equiv r/t_{\rm cool}$ is the cooling velocity.

Figure \ref{f:crit} shows the two characteristic timescales and the locus of extended H$\alpha$+[NII] filaments, for clusters (left) and groups (right). In the left panel, the blue curves represent the plasma cooling time of observed non-cool-core (NCC) clusters (\citealt{Cavagnolo:2008}) and cool-core (CC) clusters (following the recently updated constraints by \citealt{Panagoulia:2014} and \citealt{Hogan:2017}). 
The CC profiles are best fitted by a broken power law: at large radii following the cosmic adiabatic baseline (\citealt{Tozzi:2001}) with $K\propto r^{1.1}$ (for a typical 5\;keV cluster, $R_{500}\approx1.1$\;Mpc and $K_{500}\approx1600\;{\rm keV\;cm^2}$; \citealt{Sun:2009a}), while at small radii 
$K \simeq 95.4\,(r/{\rm 100\;kpc})^{2/3}\:{\rm keV\,cm^2}$.
The (X-ray) plasma cooling time is 
\begin{equation}\label{e:tcool}
t_{\rm cool}= \frac{3k_{\rm b}T}{n_{\rm e}\Lambda}, 
\end{equation}
where $n_{\rm e} = (k_{\rm b}T/K)^{3/2}$ and $\Lambda(T,Z)$ is the plasma cooling function (\citealt{Sutherland:1993}; $Z=0.3\;Z_\odot$ for clusters). The observed profiles are typically contained within 0.25\;dex (blue bands; $\approx90$\% confidence level). 
The 0.5\,-\,1\;Gyr dotted black band crudely separates the observed clusters hosting (below) or not hosting (above) extended warm filaments traced via H$\alpha$+[NII] emission (e.g.,  \citealt{Hogan:2017}).

The eddy turnover timescale, i.e., the time which a turbulent vortex requires to gyrate -- thus producing density fluctuations
in a stratified halo -- is 
\begin{equation}\label{e:teddy}
t_{\rm eddy} = 2\pi\,\frac{r^{2/3} L^{1/3}}{\sigma_{v,L}}, 
\end{equation}
where $\sigma_{v,L}$ is the velocity dispersion at the injection scale and using
the Kolmogorov cascade $\sigma_v\propto l^{1/3}$ appropriate for subsonic turbulence (\citealt{Gaspari:2013_coma}; we use here the approximation that smaller radii contain eddies of smaller size, $l \sim r$).\footnote{The eddy turnover time differs from the turbulence dissipation time, which is 10\,-\,30$\times$ longer for subsonic turbulence, $t_{\rm diss}\sim {\rm Ma_{\rm 3d}}^{-2}\, t_{\rm eddy}$. At this level, the fluctuation generation (and related condensation) outweighs the slow turbulent heating.} 
Turbulence is mainly injected via AGN feedback in the core. While it was previously difficult to assess the plasma kinematics,
the ensemble measurement makes it now trivial to apply the turbulence velocity dispersion derived from one of the condensed phases. 
We use the mean three-dimensional ($\sigma_v=\sqrt{3}\,\sigma_{v, {\rm 1d}}$) {\it ensemble} velocity dispersion, $\sigma_{v, L}\simeq 242\;\kms$,
found in the AGN feedback simulation 
and similarly in the observational sample (Fig.~\ref{f:comp}). We remark the ensemble dispersion is comparable in all phases, cold, warm, or hot.
The injection scale $L$ can be traced by the diameter of the bubble inflated by the AGN outflows/jets, which start to decelerate as they deposit the kinetic energy (or, alternatively, from the size of the warm gas nebula; more below).\footnote{
We note for the initial, brief cavity trailing stage and partial volume-filling turbulence 
(e.g., \citealt{Brighenti:2015}),
the criterion should be applied in sectors (of size $l$) rather than azimuthally.
The $C$ criterion can also be applied for the rarer merger or sloshing events, 
adopting their maximum diameter as the injection scale.
} 
From the large high-quality sample of \citet{Shin:2016}, observed 5 keV cluster bubbles have a typical diameter $L\approx25$\;kpc. Given the weak dependence on the injection scale ($\propto L^{1/3}$), the eddy time scatter for a given system is driven by $\sigma_v$, which has a 90\% confidence level of 0.2 dex (Fig.~\ref{f:comp}). 

The key result from Fig.~\ref{f:crit} is that, despite its simplicity, the crossing of $t_{\rm cool}$ and $t_{\rm eddy}$ traces the condensation region.
For our analyzed galaxy clusters (Tab.~\ref{t:ens}), the logarithmic median and dispersion of the extent radius ($\approx R_{\rm ex}$) are
$\log R/{\rm kpc} = 1.14\pm0.29$, i.e., a 2$\sigma$ range of 3.5\,-\,52.5\;kpc (which is close to that found by \citealt{McDonald:2010} in a smaller sample of CC clusters with detected warm filaments; dashed ellipse). 
This range is tracked by our proposed dimensionless cooling number $C\approx 1$ (see the overlapping blue and red bands).
Around such a threshold, turbulence in a {\it stratified} halo drives significant density/entropy perturbations in the hot phase, $\delta \rho/\rho\sim \sigma_{v,\rm{1d}}/c_{\rm s}$ (cf.~\citealt{Gaspari:2014_coma2}); the overdense hot gas rapidly cools down, promoting the multiphase cascade down to warm filaments and molecular clouds.
The cospatiality between different phases, in particular the soft X-ray and radial extent of H$\alpha$ gas, corroborates this scenario (e.g., \citealt{Hogan:2017b}).
We remark that this process differs from linear thermal instability models, which assume tiny perturbations growing exponentially against the restoring buoyancy force\footnote{The related timescale is the free-fall time (defining the TI ratio), since the buoyancy (Brunt-V\"ais\"al\"a) frequency is 
$\omega_{\rm BV}\equiv[g/(\gamma r)\;(d\ln K/d\ln r)]^{1/2} \approx t_{\rm ff}^{-1}$ 
for non-isentropic CC clusters.} in a heated halo (indeed, CCA can develop even in a non-heated atmosphere, given significant turbulent fluctuations).

An important result is that within the condensation region, both timescales do not deviate drastically from each other ($C\sim1$), except in the very inner region, implying that both generation of fluctuations and cooling act at fairly concurrent times, inducing the drop-out of warm filaments with multiple scales within the extent radius (a fraction of which triggers the central AGN). In other words, $C\sim1$ naturally traces what other studies refer to as `precipitation'-feedback balance threshold.
If the plasma cooling timescale were too short with perturbations that are weak and/or injected at very large scales ($C\ll 1$), a monolithic overcooling of the X-ray halo would develop without extended multiphase structures. This is expected to be infrequent (e.g., A2029 and A2107) because of the AGN feedback self-regulation which elevates $t_{\rm cool}$ back while decreasing $t_{\rm eddy}$ via the injection of turbulence. 
Conversely, when the cooling time is too long\footnote{If the injection scale stays the same (e.g., without any complementary cosmic-weather turbulence; \citealt{Lau:2017}) the red curve should flatten out, hence the dashed prolongation in Fig.~\ref{f:crit}. In any case, $t_{\rm cool}$ at large radii is expected to stay above this line.} ($C\gg1$),
even significant perturbations (e.g., driven by mergers) cannot induce multiphase condensation, thus preserving the NCC structure (an exemplary case is the Coma cluster). 

The right panel in Fig.~\ref{f:crit} shows the same timescales as above but for 1 keV groups/massive galaxies. Typical massive groups have $Z=1\;Z_\odot$ and $K_{500} \simeq 356\,{\rm keV\,cm^2}$ at $R_{500}\simeq 470$\;kpc (\citealt{Sun:2009a}). Cluster-like NCC groups with a flat and elevated entropy core seem nonexistent (although a more complete sample is required; see also \citealt{OSullivan:2017}). Following \citet{Sun:2009a} constraints, we plot as `NCC' the upper envelope of their observed groups. At large radii, the profiles follow the adiabatic baseline, while within $R_{500}$ the CC groups trace a similar scaling to that found by \citet{Panagoulia:2014}. Given the non-flattening entropy profiles of groups, the H$\alpha$ filament threshold is best taken within 5\,-\,15\;kpc and not as a straight line, where multiphase groups show significantly lower entropy corresponding to $t_{\rm cool}\sim$\,100\;Myr (\citealt{Werner:2014}). 
Regarding the eddy time, following the observational constraints by \citet{Shin:2016}, the driven AGN bubbles in groups show smaller average injection scales with diameters $L\approx5$\;kpc. Because of the self-regulated, thus diminished, AGN feedback power (in part counterbalanced by the smaller deposition volume), the average turbulent velocity dispersions are lower too. For our H$\alpha$+[NII] emitting groups, the median is $\approx$\,20\% lower, so we use a reference 3D velocity dispersion
$\sigma_{v,L}\simeq 190\;\kms$ 
(see also \citealt{Ogorzalek:2017}). 

The same main result for clusters applies to groups, but with the condensation region ($C\approx1$) now more compressed between radii of 0.6\,-\,15\;kpc.
For our group sample, we find a logarithmic median and dispersion for the extent radius of
$\log R/{\rm kpc}=0.44\pm0.24$,
i.e., a 2$\sigma$ range of 0.9\,-\,8.3\;kpc. \citet{McDonald:2011a} find a similar median around a few kpc, with a slightly wider range, 0.5\,-\,18\;kpc (dashed ellipse). 
Unlike clusters, some massive groups without extended multiphase filaments (e.g., NGC 4472 and NGC 1399) and residing at the CC-NCC transition can still have low cooling times (below 50 Myr) within the inner $r<500$\;pc, crossing below the eddy time. This is likely related to the ubiquitous presence in groups of central kpc-scale warm `coronae' (\citealt{Sun:2009b}). 

It is interesting to point out that the condensation radius is often comparable to the AGN bubble radius, and thus to the injection scale (again a sign of tight self-regulated feedback), for both clusters and groups. Therefore,
if an observation is lacking any evident cavity detection and if the goal is to quickly estimate the $C$ ratio, then $L\approx 2 R_{\rm ex}$ can be used as an alternative estimate for the injection scale. This keeps the $t_{\rm eddy}$ calculation solely based on the condensed gas properties. We remark that the $C$ ratio can be purely retrieved from observational data (e.g., H$\alpha$+[NII]) without requiring velocity dispersions or injection scales from simulations (though the two have been proven to be consistent; Fig.~\ref{f:comp}).

Addressing the key issues introduced in \S\ref{s:pcrit}, the $C\approx 1$ criterion is able to solve the unity problem, which should be the natural transition of physical processes.
It also does not show the one order of magnitude theoretical uncertainty of the TI ratio (which has possible threshold values $t_{\rm cool}/t_{\rm ff}\approx\;$8\,-\,70), though still retaining the $\pm 0.25$ dex intrinsic scatter due to system-to-system variations mainly tied to the cooling time.
Perhaps more important, the eddy time is observationally much easier and robust to measure compared with the challenging total masses,
as the large-scale velocity dispersion can be computed in several low-$T$ phases via the ensemble method (\S\ref{s:G17}). Indirect hydrostatic masses are required to compute $t_{\rm ff}$; however, 
solely considering turbulence and AGN feedback\footnote{ 
Geometrical deprojection biases, the AGN contamination, and lack of X-ray spectral resolution to constrain central temperature gradients are additional issues affecting the retrieval of hydrostatic masses in the core region.}, hydrostatic masses in the core can be off by a factor of several (e.g., \citealt{Gaspari:2011a}), inducing a major systematic uncertainty in the TI ratio.
On the other hand, the $C$ criterion takes advantage of the substantial noise reduction provided by the integrated spectrum measure, which is {\it directly} accessible from multiple frequency bands.

We note $t_{\rm ff}$ is a lower limit to the eddy time, as the condensed structures do not escape the cluster or group central potential. Moreover, AGN feedback self-regulates based on the halo X-ray luminosity (thus mass), so we still expect some degree of secondary correlation between these two dynamical timescales.
Finally, while the hot gas cooling time shows the largest variation between CC and NCC systems, it is important to note that the physical $C$ crossing threshold is still set by $t_{\rm eddy}$. The crossing gradually decreases from massive clusters to groups and small galaxies, from 0.5 Gyr to $\lta$\,100\;Myr, as the eddy time has a total mass trend of roughly $\propto M^{0.2}$.

In conclusion, we expect groups to display condensed structures more frequently than in clusters, but with a much more concentrated topology and with lower warm and cold gas masses. 
An excellent case study for the presence of soft X-ray turbulence/perturbations, cospatial warm filaments, and cold molecular gas is NGC 5044 (e.g., \citealt{Gastaldello:2009,David:2017}). 
Over different environments, 
the $C\simeq 0.6$\,-\,1.8 range (90\% interval) probes well the condensation region of observed systems
and thus the multiphase state of the core.
Future studies should extend the sample of clusters, groups, and especially low-mass galaxies, taking advantage of the linked kinematic properties between the hot, warm, and cold gas phases, and thus complementing the thermodynamical constraints set via high-resolution CCD imaging. \\

\vspace{+0.0cm}
\section{Conclusions} \label{s:conc}
\noindent
We probed the kinematic tracers of the top-down multiphase condensation cascade in both long-term AGN feedback simulations and high-resolution chaotic cold accretion runs. We complemented the theoretical predictions with new observational constraints of the proposed novel methodology, together with literature data. 
Our main results are summarized as follows.

\begin{itemize}
\item
In long-term (AGN feedback) and short-term (CCA feeding) simulations, we find evidence of a tight correlation in the LOS $\sigma_v$ between the hot X-ray phase and the condensed phases as {\it ensemble} (wide-aperture beam), allowing us to convert between different tracers in the UV, optical/IR, and radio bands. The RMS scatter from the linear best fit is $\approx$\,14\%. The tight kinematics is corroborated by the direct detections of {\it Hitomi} (X-ray) and SITELLE (optical) in the Perseus cluster.

\item 
As ensemble, the multiphase condensed structures display significant LOS velocity dispersion and mild bulk velocities. The pencil beam measurement (e.g., absorption against the AGN backlight) instead samples fewer and smaller condensed structures, substantially increasing the scatter due to the turbulence intermittency: the average velocity dispersion decreases following the eddy cascade, while the bulk velocity can reach values up to several 100 km\,s$^{-1}$.

\item
We presented new observational constraints for over 70 clusters and groups of the warm ($\sim$\,$10^4$\;K) and cold ($\lta 100$\;K) gas kinematics for the ensemble detection (Tab.~\ref{t:ens}), which can be used as reliable proxies for the turbulent velocities, especially for the challenging X-ray plasma (which would take an exposure of several 100\;ks per nearby object even for XARM or {\it Athena}). Comparing the lognormal distributions, the simulation predictions and observations are consistent and show a range $\sigma_{v,{\rm los}}\simeq 90$\,-\,$250\;\kms$, with mean $\approx$\,$150\;\kms$ in cluster cores.

\item
A novel diagnostic diagram of the (logarithmic) spectral line broadening versus line shift discriminates among the different kinematics and related scales. Both simulations and observations indicate that, while the ensemble points are confined in the upper-left region, the pencil-beam (small aperture, $<$ a few arcsec) detections can show a dual broad and narrow component 
sampling the chaotic large-scale gas or small-scale clouds falling toward the SMBH, respectively.

\item 
We showed that a new nonlinear multiphase condensation criterion (facilitated by the ensemble $\sigma_v$ conversion), i.e., the ratio of the cooling time and eddy turnover time $C\equiv t_{\rm cool}/t_{\rm eddy} = \sigma_v/v_{\rm cool} \approx 1$ 
(with a 90\% interval of $\pm 0.25$\;dex)
marks the condensation extent region,
as shown by the warm gas observations in the cores of groups and clusters. 
Besides solving the unity threshold problem, the $C$ ratio can be used to assess the multiphase state of the system
and is a much more robust and direct
observational quantity to measure -- via the ensemble, single spectrum method -- compared with the challenging total masses of the linear thermal instability ratio $t_{\rm cool}/t_{\rm ff}$.
\end{itemize}

\vspace{+0.2cm}
This study highlights the importance of undertaking multiwavelength campaigns (e.g., combining {\it Chandra}, XMM, ALMA, VLT, HST, {\it Magellan}, SITELLE, SOAR, IRAM, and MUSE), some of which our team is currently pursuing. 
The combination of the ensemble and pencil-beam method provides a powerful complementary diagnostic of the global and local multiphase gas kinematics, which can be optimally leveraged by the available and future IFU and spectroscopical collections of data.
At the same time, this work highlights the fairly unexplored potential of joint numerical and observational studies of multiphase gas. 
While future observations will expand the sample size, allowing more accurate statistics on the multiphase kinematic tracers, thanks also to new facilities (e.g., XARM, {\it Athena}, JWST, SKA, and CARMA2), more advanced simulations with additional physics and an upgraded dynamical range will be instrumental to further shed light on the formation and evolution of multiphase gas in galaxies, groups, and clusters of galaxies.
\\

\section*{\bf \scriptsize Acknowledgements}
\noindent
M.G. is supported by NASA through Einstein Postdoctoral Fellowship Award Number PF5-160137 issued by the Chandra X-ray Observatory Center, which is operated by the SAO for and on behalf of NASA under contract NAS8-03060. Support for this work was also provided by Chandra GO7-18121X. S.L.H. acknowledges support from the ERC for Advanced Grant Program number 339659 (MUSICOS).
J.H.-L. is supported by NSERC through the discovery grant and Canada Research Chair programs.
M.G.-M. is supported by NSERC through the NSERC Postgraduate Scholarships-Doctoral Program (PGS~D).   
N.W. is supported by the Lend\"ulet LP2016-11 grant awarded by the Hungarian Academy of Sciences.
A.C.E. acknowledges support from STFC grant ST/P00541/1.
S.E. acknowledges support from ASI-INAF I/009/10/0, NARO15 ASI-INAF I/037/12/0, and ASI 2015-046-R.0.
S.P. is supported by the Spanish Ministerio de Econom{\'i}a y Competitividad (MINECO, AYA2013-48226-C3-2-P, AYA2016-77237-C3-3-P) and the Generalitat Valenciana (GVACOMP2015-227).
\texttt{FLASH} code was in part developed by the DOE NNSA-ASC OASCR Flash Centre at the University of Chicago.
HPC resources were in part provided by the NASA/Ames HEC Program (SMD-16-7320, SMD-16-7321, SMD-16-7305; SMD-16-7251). 
We thank R.~Morganti, F.~Maccagni, M.~Voit, F.~Combes, B.~McNamara, C.~Feruglio, M.~Donahue, Y.~Cavecchi, V.~Moss, N.~Devereux, and the anonymous referee for several insightful discussions and constructive feedback, which helped to improve the manuscript.


\bibliographystyle{biblio}
\bibliography{biblio}

\providecommand{\SortNoop}[1]{}
\begin{thebibliography}{}
\makeatletter
\relax
\def\mn@urlcharsother{\let\do\@makeother \do\$\do\&\do\#\do\^\do\_\do\%\do\~}
\def\mn@doi{\begingroup\mn@urlcharsother \@ifnextchar [ {\mn@doi@}
  {\mn@doi@[]}}
\def\mn@doi@[#1]#2{\def\@tempa{#1}\ifx\@tempa\@empty \href
  {http://dx.doi.org/#2} {doi:#2}\else \href {http://dx.doi.org/#2} {#1}\fi
  \endgroup}
\def\mn@eprint#1#2{\mn@eprint@#1:#2::\@nil}
\def\mn@eprint@arXiv#1{\href {http://arxiv.org/abs/#1} {{\tt arXiv:#1}}}
\def\mn@eprint@dblp#1{\href {http://dblp.uni-trier.de/rec/bibtex/#1.xml}
  {dblp:#1}}
\def\mn@eprint@#1:#2:#3:#4\@nil{\def\@tempa {#1}\def\@tempb {#2}\def\@tempc
  {#3}\ifx \@tempc \@empty \let \@tempc \@tempb \let \@tempb \@tempa \fi \ifx
  \@tempb \@empty \def\@tempb {arXiv}\fi \@ifundefined
  {mn@eprint@\@tempb}{\@tempb:\@tempc}{\expandafter \expandafter \csname
  mn@eprint@\@tempb\endcsname \expandafter{\@tempc}}}

\bibitem[\protect\citeauthoryear{{Allison} et~al.,}{{Allison}
  et~al.}{2015}]{Allison:2015}
{Allison} J.~R.,  et~al., 2015, \mn@doi [\mnras] {10.1093/mnras/stv1532}, \href
  {http://adsabs.harvard.edu/abs/2015MNRAS.453.1249A} {453, 1249}

\bibitem[\protect\citeauthoryear{{Anderson}, {Gaspari}, {White}, {Wang}  \&
  {Dai}}{{Anderson} et~al.}{2015}]{Anderson:2015}
{Anderson} M.~E.,  {Gaspari} M.,  {White} S.~D.~M.,  {Wang} W.,   {Dai} X.,
  2015, \mn@doi [\mnras] {10.1093/mnras/stv437}, \href
  {http://adsabs.harvard.edu/abs/2015MNRAS.449.3806A} {449, 3806}

\bibitem[\protect\citeauthoryear{{Balbus} \& {Soker}}{{Balbus} \&
  {Soker}}{1989}]{Balbus:1989}
{Balbus} S.~A.,  {Soker} N.,  1989, \mn@doi [\apj] {10.1086/167521}, \href
  {http://adsabs.harvard.edu/abs/1989ApJ...341..611B} {341, 611}

\bibitem[\protect\citeauthoryear{{Barai}, {Murante}, {Borgani}, {Gaspari},
  {Granato}, {Monaco}  \& {Ragone-Figueroa}}{{Barai} et~al.}{2016}]{Barai:2016}
{Barai} P.,  {Murante} G.,  {Borgani} S.,  {Gaspari} M.,  {Granato} G.~L.,
  {Monaco} P.,   {Ragone-Figueroa} C.,  2016, \mn@doi [\mnras]
  {10.1093/mnras/stw1389}, \href
  {http://adsabs.harvard.edu/abs/2016MNRAS.461.1548B} {461, 1548}

\bibitem[\protect\citeauthoryear{{Brighenti}, {Mathews}  \& {Temi}}{{Brighenti}
  et~al.}{2015}]{Brighenti:2015}
{Brighenti} F.,  {Mathews} W.~G.,   {Temi} P.,  2015, \mn@doi [\apj]
  {10.1088/0004-637X/802/2/118}, \href
  {http://adsabs.harvard.edu/abs/2015ApJ...802..118B} {802, 118}

\bibitem[\protect\citeauthoryear{{Burkert} \& {Lin}}{{Burkert} \&
  {Lin}}{2000}]{Burkert:2000}
{Burkert} A.,  {Lin} D.~N.~C.,  2000, \mn@doi [\apj] {10.1086/308989}, \href
  {http://adsabs.harvard.edu/abs/2000ApJ...537..270B} {537, 270}

\bibitem[\protect\citeauthoryear{{Canning} et~al.,}{{Canning}
  et~al.}{2013}]{Canning:2013}
{Canning} R.~E.~A.,  et~al., 2013, \mn@doi [\mnras] {10.1093/mnras/stt1345},
  \href {http://adsabs.harvard.edu/abs/2013MNRAS.435.1108C} {435, 1108}

\bibitem[\protect\citeauthoryear{{Cavagnolo}, {Donahue}, {Voit}  \&
  {Sun}}{{Cavagnolo} et~al.}{2008}]{Cavagnolo:2008}
{Cavagnolo} K.~W.,  {Donahue} M.,  {Voit} G.~M.,   {Sun} M.,  2008, \mn@doi
  [\apjl] {10.1086/591665}, \href
  {http://adsabs.harvard.edu/abs/2008ApJ...683L.107C} {683, L107}

\bibitem[\protect\citeauthoryear{{Combes}, {Young}  \& {Bureau}}{{Combes}
  et~al.}{2007}]{Combes:2007}
{Combes} F.,  {Young} L.~M.,   {Bureau} M.,  2007, \mn@doi [\mnras]
  {10.1111/j.1365-2966.2007.11759.x}, \href
  {http://adsabs.harvard.edu/abs/2007MNRAS.377.1795C} {377, 1795}

\bibitem[\protect\citeauthoryear{{Dalgarno} \& {McCray}}{{Dalgarno} \&
  {McCray}}{1972}]{Dalgarno:1972}
{Dalgarno} A.,  {McCray} R.~A.,  1972, \mn@doi [\araa]
  {10.1146/annurev.aa.10.090172.002111}, \href
  {http://adsabs.harvard.edu/abs/1972ARA%26A..10..375D} {10, 375}

\bibitem[\protect\citeauthoryear{{David} et~al.,}{{David}
  et~al.}{2014}]{David:2014}
{David} L.~P.,  et~al., 2014, \mn@doi [\apj] {10.1088/0004-637X/792/2/94},
  \href {http://adsabs.harvard.edu/abs/2014ApJ...792...94D} {792, 94}

\bibitem[\protect\citeauthoryear{{David}, {Vrtilek}, {O'Sullivan}, {Jones},
  {Forman}  \& {Sun}}{{David} et~al.}{2017}]{David:2017}
{David} L.~P.,  {Vrtilek} J.,  {O'Sullivan} E.,  {Jones} C.,  {Forman} W.,
  {Sun} M.,  2017, \mn@doi [\apj] {10.3847/1538-4357/aa756c}, \href
  {http://adsabs.harvard.edu/abs/2017ApJ...842...84D} {842, 84}

\bibitem[\protect\citeauthoryear{{\SortNoop{D}}{de Plaa}, {Zhuravleva},
  {Werner}, {Kaastra}, {Churazov}, {Smith}, {Raassen}  \&
  {Grange}}{{\SortNoop{D}}{de Plaa} et~al.}{2012}]{dePlaa:2012}
{\SortNoop{D}}{de Plaa} J.,  {Zhuravleva} I.,  {Werner} N.,  {Kaastra} J.~S.,
  {Churazov} E.,  {Smith} R.~K.,  {Raassen} A.~J.~J.,   {Grange} Y.~G.,  2012,
  \mn@doi [\aap] {10.1051/0004-6361/201118404}, \href
  {http://adsabs.harvard.edu/abs/2012A%26A...539A..34D} {539, A34}

\bibitem[\protect\citeauthoryear{{De Grandi} et~al.,}{{De Grandi}
  et~al.}{2016}]{DeGrandi:2016}
{De Grandi} S.,  et~al., 2016, \mn@doi [\aap] {10.1051/0004-6361/201526641},
  \href {http://adsabs.harvard.edu/abs/2016A%26A...592A.154D} {592, A154}

\bibitem[\protect\citeauthoryear{{Devereux}}{{Devereux}}{2018}]{Devereux:2018}
{Devereux} N.,  2018, \mn@doi [\mnras] {10.1093/mnras/stx2537}, \href
  {http://adsabs.harvard.edu/abs/2018MNRAS.473.2930D} {473, 2930}

\bibitem[\protect\citeauthoryear{{Drissen}, {Bernier}, {Rousseau-Nepton},
  {Alarie}, {Robert}, {Joncas}, {Thibault}  \& {Grandmont}}{{Drissen}
  et~al.}{2010}]{Drissen:2010}
{Drissen} L.,  {Bernier} A.-P.,  {Rousseau-Nepton} L.,  {Alarie} A.,  {Robert}
  C.,  {Joncas} G.,  {Thibault} S.,   {Grandmont} F.,  2010, in Ground-based
  and Airborne Instrumentation for Astronomy III. p. 77350B,
  \mn@doi{10.1117/12.856470}

\bibitem[\protect\citeauthoryear{{Eckert} et~al.,}{{Eckert}
  et~al.}{2017a}]{Eckert:2017_ramP}
{Eckert} D.,  et~al., 2017a, \mn@doi [\aap] {10.1051/0004-6361/201730555},
  \href {http://adsabs.harvard.edu/abs/2017A%26A...605A..25E} {605, A25}

\bibitem[\protect\citeauthoryear{{Eckert}, {Gaspari}, {Vazza}, {Gastaldello},
  {Tramacere}, {Zimmer}, {Ettori}  \& {Paltani}}{{Eckert}
  et~al.}{2017b}]{Eckert:2017_PS}
{Eckert} D.,  {Gaspari} M.,  {Vazza} F.,  {Gastaldello} F.,  {Tramacere} A.,
  {Zimmer} S.,  {Ettori} S.,   {Paltani} S.,  2017b, \mn@doi [\apjl]
  {10.3847/2041-8213/aa7c1a}, \href
  {http://adsabs.harvard.edu/abs/2017ApJ...843L..29E} {843, L29}

\bibitem[\protect\citeauthoryear{{Ettori} et~al.,}{{Ettori}
  et~al.}{2013}]{Ettori:2013}
{Ettori} S.,  et~al., 2013, preprint, \href
  {http://adsabs.harvard.edu/abs/2013arXiv1306.2322E} {} (\mn@eprint {arXiv}
  {1306.2322})

\bibitem[\protect\citeauthoryear{{Fasano} \& {Franceschini}}{{Fasano} \&
  {Franceschini}}{1987}]{Fasano:1987}
{Fasano} G.,  {Franceschini} A.,  1987, \mn@doi [\mnras]
  {10.1093/mnras/225.1.155}, \href
  {http://adsabs.harvard.edu/abs/1987MNRAS.225..155F} {225, 155}

\bibitem[\protect\citeauthoryear{{Field}}{{Field}}{1965}]{Field:1965}
{Field} G.~B.,  1965, \mn@doi [\apj] {10.1086/148317}, \href
  {http://adsabs.harvard.edu/abs/1965ApJ...142..531F} {142, 531}

\bibitem[\protect\citeauthoryear{{Fisher} et~al.,}{{Fisher}
  et~al.}{2008}]{Fisher:2008}
{Fisher} R.~T.,  et~al., 2008, IBM J. Res. \& Dev., 52, 127

\bibitem[\protect\citeauthoryear{{Gaspari}}{{Gaspari}}{2015}]{Gaspari:2015_xspec}
{Gaspari} M.,  2015, \mn@doi [\mnras] {10.1093/mnrasl/slv067}, \href
  {http://adsabs.harvard.edu/abs/2015MNRAS.451L..60G} {451, L60}

\bibitem[\protect\citeauthoryear{{Gaspari}}{{Gaspari}}{2016}]{Gaspari:2016}
{Gaspari} M.,  2016, in {Kaviraj} S.,  ed.,  IAU Symposium Vol. 319, Galaxies
  at High Redshift and Their Evolution Over Cosmic Time. pp 17--20,
  \mn@doi{10.1017/S1743921315010455}

\bibitem[\protect\citeauthoryear{{Gaspari} \& {Churazov}}{{Gaspari} \&
  {Churazov}}{2013}]{Gaspari:2013_coma}
{Gaspari} M.,  {Churazov} E.,  2013, \mn@doi [\aap]
  {10.1051/0004-6361/201322295}, \href
  {http://adsabs.harvard.edu/abs/2013A%26A...559A..78G} {559, A78}

\bibitem[\protect\citeauthoryear{{Gaspari} \& {S{\c a}dowski}}{{Gaspari} \&
  {S{\c a}dowski}}{2017}]{Gaspari:2017_uni}
{Gaspari} M.,  {S{\c a}dowski} A.,  2017, \mn@doi [\apj]
  {10.3847/1538-4357/aa61a3}, \href
  {http://adsabs.harvard.edu/abs/2017ApJ...837..149G} {837, 149}

\bibitem[\protect\citeauthoryear{{Gaspari}, {{\SortNoop{m2011a}}Melioli},
  {Brighenti}  \& {D'Ercole}}{{Gaspari} et~al.}{2011a}]{Gaspari:2011a}
{Gaspari} M.,  {{\SortNoop{m2011a}}Melioli} C.,  {Brighenti} F.,   {D'Ercole}
  A.,  2011a, \mn@doi [\mnras] {10.1111/j.1365-2966.2010.17688.x}, \href
  {http://adsabs.harvard.edu/abs/2011MNRAS.411..349G} {411, 349}

\bibitem[\protect\citeauthoryear{{Gaspari}, {{\SortNoop{m2011b}}Brighenti},
  {D'Ercole}  \& {Melioli}}{{Gaspari} et~al.}{2011b}]{Gaspari:2011b}
{Gaspari} M.,  {{\SortNoop{m2011b}}Brighenti} F.,  {D'Ercole} A.,   {Melioli}
  C.,  2011b, \mn@doi [\mnras] {10.1111/j.1365-2966.2011.18806.x}, \href
  {http://adsabs.harvard.edu/abs/2011MNRAS.415.1549G} {415, 1549}

\bibitem[\protect\citeauthoryear{{Gaspari}, {\SortNoop{b}}{Brighenti}  \&
  {Temi}}{{Gaspari} et~al.}{2012a}]{Gaspari:2012b}
{Gaspari} M.,  {\SortNoop{b}}{Brighenti} F.,   {Temi} P.,  2012a, \mn@doi
  [\mnras] {10.1111/j.1365-2966.2012.21183.x}, \href
  {http://adsabs.harvard.edu/abs/2012MNRAS.424..190G} {424, 190}

\bibitem[\protect\citeauthoryear{{Gaspari}, {\SortNoop{a}}{Ruszkowski}  \&
  {Sharma}}{{Gaspari} et~al.}{2012b}]{Gaspari:2012a}
{Gaspari} M.,  {\SortNoop{a}}{Ruszkowski} M.,   {Sharma} P.,  2012b, \mn@doi
  [\apj] {10.1088/0004-637X/746/1/94}, \href
  {http://adsabs.harvard.edu/abs/2012ApJ...746...94G} {746, 94}

\bibitem[\protect\citeauthoryear{{Gaspari}, {Brighenti}  \&
  {Ruszkowski}}{{Gaspari} et~al.}{2013a}]{Gaspari:2013_rev}
{Gaspari} M.,  {Brighenti} F.,   {Ruszkowski} M.,  2013a, \mn@doi [AN]
  {10.1002/asna.201211865}, \href
  {http://adsabs.harvard.edu/abs/2013AN....334..394G} {334, 394}

\bibitem[\protect\citeauthoryear{{Gaspari}, {Ruszkowski}  \& {Oh}}{{Gaspari}
  et~al.}{2013b}]{Gaspari:2013_cca}
{Gaspari} M.,  {Ruszkowski} M.,   {Oh} S.~P.,  2013b, \mn@doi [\mnras]
  {10.1093/mnras/stt692}, \href
  {http://adsabs.harvard.edu/abs/2013MNRAS.432.3401G} {432, 3401}

\bibitem[\protect\citeauthoryear{{Gaspari}, {Churazov}, {Nagai}, {Lau}  \&
  {Zhuravleva}}{{Gaspari} et~al.}{2014}]{Gaspari:2014_coma2}
{Gaspari} M.,  {Churazov} E.,  {Nagai} D.,  {Lau} E.~T.,   {Zhuravleva} I.,
  2014, \mn@doi [\aap] {10.1051/0004-6361/201424043}, \href
  {http://adsabs.harvard.edu/abs/2014A%26A...569A..67G} {569, A67}

\bibitem[\protect\citeauthoryear{{Gaspari}, {Brighenti}  \& {Temi}}{{Gaspari}
  et~al.}{2015}]{Gaspari:2015_cca}
{Gaspari} M.,  {Brighenti} F.,   {Temi} P.,  2015, \mn@doi [\aap]
  {10.1051/0004-6361/201526151}, \href
  {http://adsabs.harvard.edu/abs/2015A%26A...579A..62G} {579, A62}

\bibitem[\protect\citeauthoryear{{Gaspari}, {Temi}  \& {Brighenti}}{{Gaspari}
  et~al.}{2017}]{Gaspari:2017_cca}
{Gaspari} M.,  {Temi} P.,   {Brighenti} F.,  2017, \mn@doi [\mnras]
  {10.1093/mnras/stw3108}, \href
  {http://adsabs.harvard.edu/abs/2017MNRAS.466..677G} {466, 677}

\bibitem[\protect\citeauthoryear{{Gastaldello}, {Buote}, {Temi}, {Brighenti},
  {Mathews}  \& {Ettori}}{{Gastaldello} et~al.}{2009}]{Gastaldello:2009}
{Gastaldello} F.,  {Buote} D.~A.,  {Temi} P.,  {Brighenti} F.,  {Mathews}
  W.~G.,   {Ettori} S.,  2009, \mn@doi [\apj] {10.1088/0004-637X/693/1/43},
  \href {http://adsabs.harvard.edu/abs/2009ApJ...693...43G} {693, 43}

\bibitem[\protect\citeauthoryear{{Hamer} et~al.,}{{Hamer}
  et~al.}{2016}]{Hamer:2016}
{Hamer} S.~L.,  et~al., 2016, \mn@doi [\mnras] {10.1093/mnras/stw1054}, \href
  {http://adsabs.harvard.edu/abs/2016MNRAS.460.1758H} {460, 1758}

\bibitem[\protect\citeauthoryear{{Hillel} \& {Soker}}{{Hillel} \&
  {Soker}}{2017}]{Hillel:2017}
{Hillel} S.,  {Soker} N.,  2017, \mn@doi [\apj] {10.3847/1538-4357/aa81c5},
  \href {http://adsabs.harvard.edu/abs/2017ApJ...845...91H} {845, 91}

\bibitem[\protect\citeauthoryear{{Hitomi Collaboration}}{{Hitomi
  Collaboration}}{2016}]{Hitomi:2016}
{Hitomi Collaboration} 2016, \mn@doi [\nat] {10.1038/nature18627}, \href
  {http://adsabs.harvard.edu/abs/2016Natur.535..117H} {535, 117}

\bibitem[\protect\citeauthoryear{{Hitomi Collaboration} et~al.,}{{Hitomi
  Collaboration} et~al.}{2017}]{Hitomi:2017}
{Hitomi Collaboration} et~al., 2017, preprint, \href
  {http://adsabs.harvard.edu/abs/2017arXiv171100240H} {} (\mn@eprint {arXiv}
  {1711.00240})

\bibitem[\protect\citeauthoryear{{Hofmann}, {Sanders}, {Nandra}, {Clerc}  \&
  {Gaspari}}{{Hofmann} et~al.}{2016}]{Hofmann:2016}
{Hofmann} F.,  {Sanders} J.~S.,  {Nandra} K.,  {Clerc} N.,   {Gaspari} M.,
  2016, \mn@doi [\aap] {10.1051/0004-6361/201526925}, \href
  {http://adsabs.harvard.edu/abs/2016A%26A...585A.130H} {585, A130}

\bibitem[\protect\citeauthoryear{{Hogan}}{{Hogan}}{2014}]{Hogan:2014}
{Hogan} M.~T.,  2014, PhD thesis, Durham University

\bibitem[\protect\citeauthoryear{{Hogan}, {McNamara}, {Pulido}, {Nulsen},
  {Russell}, {Vantyghem}, {Edge}  \& {Main}}{{Hogan}
  et~al.}{2017a}]{Hogan:2017b}
{Hogan} M.~T.,  {McNamara} B.~R.,  {Pulido} F.,  {Nulsen} P.~E.~J.,  {Russell}
  H.~R.,  {Vantyghem} A.~N.,  {Edge} A.~C.,   {Main} R.~A.,  2017a, \mn@doi
  [\apj] {10.3847/1538-4357/aa5f56}, \href
  {http://adsabs.harvard.edu/abs/2017ApJ...837...51H} {837, 51}

\bibitem[\protect\citeauthoryear{{Hogan} et~al.,}{{Hogan}
  et~al.}{2017b}]{Hogan:2017}
{Hogan} M.~T.,  et~al., 2017b, \mn@doi [\apj] {10.3847/1538-4357/aa9af3}, \href
  {http://adsabs.harvard.edu/abs/2017ApJ...851...66H} {851, 66}

\bibitem[\protect\citeauthoryear{{Huber}, {Tchernin}, {Eckert}, {Farnier},
  {Manalaysay}, {Straumann}  \& {Walter}}{{Huber} et~al.}{2013}]{Huber:2013}
{Huber} B.,  {Tchernin} C.,  {Eckert} D.,  {Farnier} C.,  {Manalaysay} A.,
  {Straumann} U.,   {Walter} R.,  2013, \mn@doi [\aap]
  {10.1051/0004-6361/201321947}, \href
  {http://adsabs.harvard.edu/abs/2013A%26A...560A..64H} {560, A64}

\bibitem[\protect\citeauthoryear{{Khatri} \& {Gaspari}}{{Khatri} \&
  {Gaspari}}{2016}]{Khatri:2016}
{Khatri} R.,  {Gaspari} M.,  2016, \mn@doi [\mnras] {10.1093/mnras/stw2027},
  \href {http://adsabs.harvard.edu/abs/2016MNRAS.463..655K} {463, 655}

\bibitem[\protect\citeauthoryear{{Kim}, {Ostriker}  \& {Kim}}{{Kim}
  et~al.}{2013}]{Kim:2013}
{Kim} C.-G.,  {Ostriker} E.~C.,   {Kim} W.-T.,  2013, \mn@doi [\apj]
  {10.1088/0004-637X/776/1/1}, \href
  {http://adsabs.harvard.edu/abs/2013ApJ...776....1K} {776, 1}

\bibitem[\protect\citeauthoryear{{Komarov}, {Churazov}, {Kunz}  \&
  {Schekochihin}}{{Komarov} et~al.}{2016}]{Komarov:2016}
{Komarov} S.~V.,  {Churazov} E.~M.,  {Kunz} M.~W.,   {Schekochihin} A.~A.,
  2016, \mn@doi [\mnras] {10.1093/mnras/stw963}, \href
  {http://adsabs.harvard.edu/abs/2016MNRAS.460..467K} {460, 467}

\bibitem[\protect\citeauthoryear{{Koyama} \& {Inutsuka}}{{Koyama} \&
  {Inutsuka}}{2000}]{Koyama:2000}
{Koyama} H.,  {Inutsuka} S.-I.,  2000, \mn@doi [\apj] {10.1086/308594}, \href
  {http://adsabs.harvard.edu/abs/2000ApJ...532..980K} {532, 980}

\bibitem[\protect\citeauthoryear{{Kumar}, {Eichler}, {Gaspari}  \&
  {Spitkovsky}}{{Kumar} et~al.}{2017}]{Kumar:2017}
{Kumar} R.,  {Eichler} D.,  {Gaspari} M.,   {Spitkovsky} A.,  2017, \mn@doi
  [\apj] {10.3847/1538-4357/835/2/295}, \href
  {http://adsabs.harvard.edu/abs/2017ApJ...835..295K} {835, 295}

\bibitem[\protect\citeauthoryear{{Lau}, {Kravtsov}  \& {Nagai}}{{Lau}
  et~al.}{2009}]{Lau:2009}
{Lau} E.~T.,  {Kravtsov} A.~V.,   {Nagai} D.,  2009, \mn@doi [\apj]
  {10.1088/0004-637X/705/2/1129}, \href
  {http://adsabs.harvard.edu/abs/2009ApJ...705.1129L} {705, 1129}

\bibitem[\protect\citeauthoryear{{Lau}, {Gaspari}, {Nagai}  \& {Coppi}}{{Lau}
  et~al.}{2017}]{Lau:2017}
{Lau} E.~T.,  {Gaspari} M.,  {Nagai} D.,   {Coppi} P.,  2017, \mn@doi [\apj]
  {10.3847/1538-4357/aa8c00}, \href
  {http://adsabs.harvard.edu/abs/2017ApJ...849...54L} {849, 54}

\bibitem[\protect\citeauthoryear{{Li} \& {Bryan}}{{Li} \&
  {Bryan}}{2014}]{Li:2014}
{Li} Y.,  {Bryan} G.~L.,  2014, \mn@doi [\apj] {10.1088/0004-637X/789/1/54},
  \href {http://adsabs.harvard.edu/abs/2014ApJ...789...54L} {789, 54}

\bibitem[\protect\citeauthoryear{{Loewenstein} \& {Fabian}}{{Loewenstein} \&
  {Fabian}}{1990}]{Loewenstein:1990}
{Loewenstein} M.,  {Fabian} A.~C.,  1990, \mn@doi [\mnras]
  {10.1093/mnras/242.2.120}, \href
  {http://adsabs.harvard.edu/abs/1990MNRAS.242..120L} {242, 120}

\bibitem[\protect\citeauthoryear{{Maccagni}, {Morganti}, {Oosterloo},
  {Ger{\'e}b}  \& {Maddox}}{{Maccagni} et~al.}{2017}]{Maccagni:2017}
{Maccagni} F.~M.,  {Morganti} R.,  {Oosterloo} T.~A.,  {Ger{\'e}b} K.,
  {Maddox} N.,  2017, \mn@doi [\aap] {10.1051/0004-6361/201730563}, \href
  {http://adsabs.harvard.edu/abs/2017A%26A...604A..43M} {604, A43}

\bibitem[\protect\citeauthoryear{{Martin}, {Drissen}  \& {Joncas}}{{Martin}
  et~al.}{2015}]{Martin:2015}
{Martin} T.,  {Drissen} L.,   {Joncas} G.,  2015, in {Taylor} A.~R.,
  {Rosolowsky} E.,  eds,  Astronomical Society of the Pacific Conference Series
  Vol. 495, Astronomical Data Analysis Software an Systems XXIV (ADASS XXIV).
  p.~327

\bibitem[\protect\citeauthoryear{{McCourt}, {Sharma}, {Quataert}  \&
  {Parrish}}{{McCourt} et~al.}{2012}]{McCourt:2012}
{McCourt} M.,  {Sharma} P.,  {Quataert} E.,   {Parrish} I.~J.,  2012, \mn@doi
  [\mnras] {10.1111/j.1365-2966.2011.19972.x}, \href
  {http://adsabs.harvard.edu/abs/2012MNRAS.419.3319M} {419, 3319}

\bibitem[\protect\citeauthoryear{{McDonald}, {Veilleux}, {Rupke}  \&
  {Mushotzky}}{{McDonald} et~al.}{2010}]{McDonald:2010}
{McDonald} M.,  {Veilleux} S.,  {Rupke} D.~S.~N.,   {Mushotzky} R.,  2010,
  \mn@doi [\apj] {10.1088/0004-637X/721/2/1262}, \href
  {http://adsabs.harvard.edu/abs/2010ApJ...721.1262M} {721, 1262}

\bibitem[\protect\citeauthoryear{{McDonald}, {Veilleux}  \&
  {Mushotzky}}{{McDonald} et~al.}{2011}]{McDonald:2011a}
{McDonald} M.,  {Veilleux} S.,   {Mushotzky} R.,  2011, \mn@doi [\apj]
  {10.1088/0004-637X/731/1/33}, \href
  {http://adsabs.harvard.edu/abs/2011ApJ...731...33M} {731, 33}

\bibitem[\protect\citeauthoryear{{McDonald}, {Veilleux}  \& {Rupke}}{{McDonald}
  et~al.}{2012}]{McDonald:2012_Ha}
{McDonald} M.,  {Veilleux} S.,   {Rupke} D.~S.~N.,  2012, \mn@doi [\apj]
  {10.1088/0004-637X/746/2/153}, \href
  {http://adsabs.harvard.edu/abs/2012ApJ...746..153M} {746, 153}

\bibitem[\protect\citeauthoryear{{McKee} \& {Ostriker}}{{McKee} \&
  {Ostriker}}{1977}]{McKee:1977}
{McKee} C.~F.,  {Ostriker} J.~P.,  1977, \mn@doi [\apj] {10.1086/155667}, \href
  {http://adsabs.harvard.edu/abs/1977ApJ...218..148M} {218, 148}

\bibitem[\protect\citeauthoryear{{McNamara} \& {Nulsen}}{{McNamara} \&
  {Nulsen}}{2012}]{McNamara:2012}
{McNamara} B.~R.,  {Nulsen} P.~E.~J.,  2012, \mn@doi [New J. Phys.]
  {10.1088/1367-2630/14/5/055023}, \href
  {http://adsabs.harvard.edu/abs/2012NJPh...14e5023M} {14, 055023}

\bibitem[\protect\citeauthoryear{{McNamara} et~al.,}{{McNamara}
  et~al.}{2014}]{McNamara:2014}
{McNamara} B.~R.,  et~al., 2014, \mn@doi [\apj] {10.1088/0004-637X/785/1/44},
  \href {http://adsabs.harvard.edu/abs/2014ApJ...785...44M} {785, 44}

\bibitem[\protect\citeauthoryear{{McNamara}, {Russell}, {Nulsen}, {Hogan},
  {Fabian}, {Pulido}  \& {Edge}}{{McNamara} et~al.}{2016}]{McNamara:2016}
{McNamara} B.~R.,  {Russell} H.~R.,  {Nulsen} P.~E.~J.,  {Hogan} M.~T.,
  {Fabian} A.~C.,  {Pulido} F.,   {Edge} A.~C.,  2016, \mn@doi [\apj]
  {10.3847/0004-637X/830/2/79}, \href
  {http://adsabs.harvard.edu/abs/2016ApJ...830...79M} {830, 79}

\bibitem[\protect\citeauthoryear{{Meece}, {Voit}  \& {O'Shea}}{{Meece}
  et~al.}{2017}]{Meece:2017}
{Meece} G.~R.,  {Voit} G.~M.,   {O'Shea} B.~W.,  2017, \mn@doi [\apj]
  {10.3847/1538-4357/aa6fb1}, \href
  {http://adsabs.harvard.edu/abs/2017ApJ...841..133M} {841, 133}

\bibitem[\protect\citeauthoryear{{Miniati}}{{Miniati}}{2014}]{Miniati:2014}
{Miniati} F.,  2014, \mn@doi [\apj] {10.1088/0004-637X/782/1/21}, \href
  {http://adsabs.harvard.edu/abs/2014ApJ...782...21M} {782, 21}

\bibitem[\protect\citeauthoryear{{Moss} et~al.,}{{Moss}
  et~al.}{2017}]{Moss:2017}
{Moss} V.~A.,  et~al., 2017, \mn@doi [\mnras] {10.1093/mnras/stx1679}, \href
  {http://adsabs.harvard.edu/abs/2017MNRAS.471.2952M} {471, 2952}

\bibitem[\protect\citeauthoryear{{Nagai}, {Lau}, {Avestruz}, {Nelson}  \&
  {Rudd}}{{Nagai} et~al.}{2013}]{Nagai:2013}
{Nagai} D.,  {Lau} E.~T.,  {Avestruz} C.,  {Nelson} K.,   {Rudd} D.~H.,  2013,
  \mn@doi [\apj] {10.1088/0004-637X/777/2/137}, \href
  {http://adsabs.harvard.edu/abs/2013ApJ...777..137N} {777, 137}

\bibitem[\protect\citeauthoryear{{O'Sullivan} et~al.,}{{O'Sullivan}
  et~al.}{2017}]{OSullivan:2017}
{O'Sullivan} E.,  et~al., 2017, \mn@doi [\mnras] {10.1093/mnras/stx2078}, \href
  {http://adsabs.harvard.edu/abs/2017MNRAS.472.1482O} {472, 1482}

\bibitem[\protect\citeauthoryear{{Ogorzalek} et~al.,}{{Ogorzalek}
  et~al.}{2017}]{Ogorzalek:2017}
{Ogorzalek} A.,  et~al., 2017, \mn@doi [\mnras] {10.1093/mnras/stx2030}, \href
  {http://adsabs.harvard.edu/abs/2017MNRAS.472.1659O} {472, 1659}

\bibitem[\protect\citeauthoryear{{Paczy{\'n}ski} \& {Wiita}}{{Paczy{\'n}ski} \&
  {Wiita}}{1980}]{Paczynski:1980}
{Paczy{\'n}ski} B.,  {Wiita} P.~J.,  1980, \aap, \href
  {http://adsabs.harvard.edu/abs/1980A%26A....88...23P} {88, 23}

\bibitem[\protect\citeauthoryear{{Panagoulia}, {Fabian}  \&
  {Sanders}}{{Panagoulia} et~al.}{2014}]{Panagoulia:2014}
{Panagoulia} E.~K.,  {Fabian} A.~C.,   {Sanders} J.~S.,  2014, \mn@doi [\mnras]
  {10.1093/mnras/stt2349}, \href
  {http://adsabs.harvard.edu/abs/2014MNRAS.438.2341P} {438, 2341}

\bibitem[\protect\citeauthoryear{{Pinto} et~al.,}{{Pinto}
  et~al.}{2015}]{Pinto:2015}
{Pinto} C.,  et~al., 2015, \mn@doi [\aap] {10.1051/0004-6361/201425278}, \href
  {http://adsabs.harvard.edu/abs/2015A%26A...575A..38P} {575, A38}

\bibitem[\protect\citeauthoryear{{Pizzolato} \& {Soker}}{{Pizzolato} \&
  {Soker}}{2005}]{Pizzolato:2005}
{Pizzolato} F.,  {Soker} N.,  2005, \mn@doi [\apj] {10.1086/444344}, \href
  {http://adsabs.harvard.edu/abs/2005ApJ...632..821P} {632, 821}

\bibitem[\protect\citeauthoryear{{Press}, {Teukolsky}, {Vetterling}  \&
  {Flannery}}{{Press} et~al.}{1992}]{NRec:1992}
{Press} W.~H.,  {Teukolsky} S.~A.,  {Vetterling} W.~T.,   {Flannery} B.~P.,
  1992, {Numerical recipes in FORTRAN. The art of scientific computing,
  Cambridge: University Press}

\bibitem[\protect\citeauthoryear{{Pulido} et~al.,}{{Pulido}
  et~al.}{2017}]{Pulido:2017}
{Pulido} F.~A.,  et~al., 2017, preprint, \href
  {http://adsabs.harvard.edu/abs/2017arXiv171004664P} {} (\mn@eprint {arXiv}
  {1710.04664})

\bibitem[\protect\citeauthoryear{{Russell} et~al.,}{{Russell}
  et~al.}{2016}]{Russell:2016}
{Russell} H.~R.,  et~al., 2016, \mn@doi [\mnras] {10.1093/mnras/stw409}, \href
  {http://adsabs.harvard.edu/abs/2016MNRAS.458.3134R} {458, 3134}

\bibitem[\protect\citeauthoryear{{Salom{\'e}} et~al.,}{{Salom{\'e}}
  et~al.}{2006}]{Salome:2006}
{Salom{\'e}} P.,  et~al., 2006, \mn@doi [\aap] {10.1051/0004-6361:20054745},
  \href {http://adsabs.harvard.edu/abs/2006A%26A...454..437S} {454, 437}

\bibitem[\protect\citeauthoryear{{Salom{\'e}}, {Combes}, {Revaz}, {Edge},
  {Hatch}, {Fabian}  \& {Johnstone}}{{Salom{\'e}} et~al.}{2008}]{Salome:2008}
{Salom{\'e}} P.,  {Combes} F.,  {Revaz} Y.,  {Edge} A.~C.,  {Hatch} N.~A.,
  {Fabian} A.~C.,   {Johnstone} R.~M.,  2008, \mn@doi [\aap]
  {10.1051/0004-6361:200809493}, \href
  {http://adsabs.harvard.edu/abs/2008A%26A...484..317S} {484, 317}

\bibitem[\protect\citeauthoryear{{Sanders} \& {Fabian}}{{Sanders} \&
  {Fabian}}{2013}]{Sanders:2013}
{Sanders} J.~S.,  {Fabian} A.~C.,  2013, \mn@doi [\mnras]
  {10.1093/mnras/sts543}, \href
  {http://adsabs.harvard.edu/abs/2013MNRAS.429.2727S} {429, 2727}

\bibitem[\protect\citeauthoryear{{Sharma}, {Parrish}  \& {Quataert}}{{Sharma}
  et~al.}{2010}]{Sharma:2010}
{Sharma} P.,  {Parrish} I.~J.,   {Quataert} E.,  2010, \mn@doi [\apj]
  {10.1088/0004-637X/720/1/652}, \href
  {http://adsabs.harvard.edu/abs/2010ApJ...720..652S} {720, 652}

\bibitem[\protect\citeauthoryear{{Sharma}, {McCourt}, {Quataert}  \&
  {Parrish}}{{Sharma} et~al.}{2012}]{Sharma:2012}
{Sharma} P.,  {McCourt} M.,  {Quataert} E.,   {Parrish} I.~J.,  2012, \mn@doi
  [\mnras] {10.1111/j.1365-2966.2011.20246.x}, \href
  {http://adsabs.harvard.edu/abs/2012MNRAS.420.3174S} {420, 3174}

\bibitem[\protect\citeauthoryear{{Shin}, {Woo}  \& {Mulchaey}}{{Shin}
  et~al.}{2016}]{Shin:2016}
{Shin} J.,  {Woo} J.-H.,   {Mulchaey} J.~S.,  2016, \mn@doi [\apjs]
  {10.3847/1538-4365/227/2/31}, \href
  {http://adsabs.harvard.edu/abs/2016ApJS..227...31S} {227, 31}

\bibitem[\protect\citeauthoryear{{Smith} \& {Edge}}{{Smith} \&
  {Edge}}{2017}]{Smith:2017}
{Smith} R.~J.,  {Edge} A.~C.,  2017, \mn@doi [\mnras] {10.1093/mnrasl/slx107},
  \href {http://adsabs.harvard.edu/abs/2017MNRAS.471L..66S} {471, L66}

\bibitem[\protect\citeauthoryear{{Soker}}{{Soker}}{2016}]{Soker:2016}
{Soker} N.,  2016, \mn@doi [\nar] {10.1016/j.newar.2016.08.002}, \href
  {http://adsabs.harvard.edu/abs/2016NewAR..75....1S} {75, 1}

\bibitem[\protect\citeauthoryear{{Sun}}{{Sun}}{2009}]{Sun:2009b}
{Sun} M.,  2009, \mn@doi [\apj] {10.1088/0004-637X/704/2/1586}, \href
  {http://adsabs.harvard.edu/abs/2009ApJ...704.1586S} {704, 1586}

\bibitem[\protect\citeauthoryear{{Sun}}{{Sun}}{2012}]{Sun:2012}
{Sun} M.,  2012, \mn@doi [New Journal of Physics]
  {10.1088/1367-2630/14/4/045004}, \href
  {http://adsabs.harvard.edu/abs/2012NJPh...14d5004S} {14, 045004}

\bibitem[\protect\citeauthoryear{{Sun}, {Voit}, {Donahue}, {Jones}, {Forman}
  \& {Vikhlinin}}{{Sun} et~al.}{2009}]{Sun:2009a}
{Sun} M.,  {Voit} G.~M.,  {Donahue} M.,  {Jones} C.,  {Forman} W.,
  {Vikhlinin} A.,  2009, \mn@doi [\apj] {10.1088/0004-637X/693/2/1142}, \href
  {http://adsabs.harvard.edu/abs/2009ApJ...693.1142S} {693, 1142}

\bibitem[\protect\citeauthoryear{{Sutherland} \& {Dopita}}{{Sutherland} \&
  {Dopita}}{1993}]{Sutherland:1993}
{Sutherland} R.~S.,  {Dopita} M.~A.,  1993, \mn@doi [\apjs] {10.1086/191823},
  \href {http://adsabs.harvard.edu/abs/1993ApJS...88..253S} {88, 253}

\bibitem[\protect\citeauthoryear{{Temi}, {Amblard}, {Gitti}, {Brighenti},
  {Gaspari}, {Mathews}  \& {David}}{{Temi} et~al.}{2017}]{Temi:2017}
{Temi} P.,  {Amblard} A.,  {Gitti} M.,  {Brighenti} F.,  {Gaspari} M.,
  {Mathews} W.~G.,   {David} L.,  2017, preprint, \href
  {http://adsabs.harvard.edu/abs/2017arXiv171110630T} {} (\mn@eprint {arXiv}
  {1711.10630})

\bibitem[\protect\citeauthoryear{{Tombesi}, {Cappi}, {Reeves}, {Nemmen},
  {Braito}, {Gaspari}  \& {Reynolds}}{{Tombesi} et~al.}{2013}]{Tombesi:2013}
{Tombesi} F.,  {Cappi} M.,  {Reeves} J.~N.,  {Nemmen} R.~S.,  {Braito} V.,
  {Gaspari} M.,   {Reynolds} C.~S.,  2013, \mn@doi [\mnras]
  {10.1093/mnras/sts692}, \href
  {http://adsabs.harvard.edu/abs/2013MNRAS.430.1102T} {430, 1102}

\bibitem[\protect\citeauthoryear{{Tozzi} \& {Norman}}{{Tozzi} \&
  {Norman}}{2001}]{Tozzi:2001}
{Tozzi} P.,  {Norman} C.,  2001, \mn@doi [\apj] {10.1086/318237}, \href
  {http://adsabs.harvard.edu/abs/2001ApJ...546...63T} {546, 63}

\bibitem[\protect\citeauthoryear{{Tremblay} et~al.,}{{Tremblay}
  et~al.}{2015}]{Tremblay:2015}
{Tremblay} G.~R.,  et~al., 2015, \mn@doi [\mnras] {10.1093/mnras/stv1151},
  \href {http://adsabs.harvard.edu/abs/2015MNRAS.451.3768T} {451, 3768}

\bibitem[\protect\citeauthoryear{{Tremblay} et~al.,}{{Tremblay}
  et~al.}{2016}]{Tremblay:2016}
{Tremblay} G.~R.,  et~al., 2016, \mn@doi [\nat] {10.1038/nature17969}, \href
  {http://adsabs.harvard.edu/abs/2016Natur.534..218T} {534, 218}

\bibitem[\protect\citeauthoryear{{Valentini} \& {Brighenti}}{{Valentini} \&
  {Brighenti}}{2015}]{Valentini:2015}
{Valentini} M.,  {Brighenti} F.,  2015, \mn@doi [\mnras]
  {10.1093/mnras/stv090}, \href
  {http://adsabs.harvard.edu/abs/2015MNRAS.448.1979V} {448, 1979}

\bibitem[\protect\citeauthoryear{{Vantyghem} et~al.,}{{Vantyghem}
  et~al.}{2016}]{Vantyghem:2016}
{Vantyghem} A.~N.,  et~al., 2016, \mn@doi [\apj] {10.3847/0004-637X/832/2/148},
  \href {http://adsabs.harvard.edu/abs/2016ApJ...832..148V} {832, 148}

\bibitem[\protect\citeauthoryear{{Vazza}, {Brunetti}, {Gheller}, {Brunino}  \&
  {Br{\"u}ggen}}{{Vazza} et~al.}{2011}]{Vazza:2011}
{Vazza} F.,  {Brunetti} G.,  {Gheller} C.,  {Brunino} R.,   {Br{\"u}ggen} M.,
  2011, \mn@doi [\aap] {10.1051/0004-6361/201016015}, \href
  {http://adsabs.harvard.edu/abs/2011A%26A...529A..17V} {529, A17}

\bibitem[\protect\citeauthoryear{{Vermeulen} et~al.,}{{Vermeulen}
  et~al.}{2003}]{Vermeulen:2003}
{Vermeulen} R.~C.,  et~al., 2003, \mn@doi [\aap] {10.1051/0004-6361:20030468},
  \href {http://adsabs.harvard.edu/abs/2003A%26A...404..861V} {404, 861}

\bibitem[\protect\citeauthoryear{{Voit}, {Donahue}, {Bryan}  \&
  {McDonald}}{{Voit} et~al.}{2015a}]{Voit:2015_nat}
{Voit} G.~M.,  {Donahue} M.,  {Bryan} G.~L.,   {McDonald} M.,  2015a, \mn@doi
  [\nat] {10.1038/nature14167}, \href
  {http://adsabs.harvard.edu/abs/2015Natur.519..203V} {519, 203}

\bibitem[\protect\citeauthoryear{{Voit}, {Donahue}, {O'Shea}, {Bryan}, {Sun}
  \& {Werner}}{{Voit} et~al.}{2015b}]{Voit:2015_gE}
{Voit} G.~M.,  {Donahue} M.,  {O'Shea} B.~W.,  {Bryan} G.~L.,  {Sun} M.,
  {Werner} N.,  2015b, \mn@doi [\apjl] {10.1088/2041-8205/803/2/L21}, \href
  {http://adsabs.harvard.edu/abs/2015ApJ...803L..21V} {803, L21}

\bibitem[\protect\citeauthoryear{{Voit}, {Meece}, {Li}, {O'Shea}, {Bryan}  \&
  {Donahue}}{{Voit} et~al.}{2017}]{Voit:2017}
{Voit} G.~M.,  {Meece} G.,  {Li} Y.,  {O'Shea} B.~W.,  {Bryan} G.~L.,
  {Donahue} M.,  2017, \mn@doi [\apj] {10.3847/1538-4357/aa7d04}, \href
  {http://adsabs.harvard.edu/abs/2017ApJ...845...80V} {845, 80}

\bibitem[\protect\citeauthoryear{{Werner}, {Zhuravleva}, {Churazov},
  {Simionescu}, {Allen}, {Forman}, {Jones}  \& {Kaastra}}{{Werner}
  et~al.}{2009}]{Werner:2009}
{Werner} N.,  {Zhuravleva} I.,  {Churazov} E.,  {Simionescu} A.,  {Allen}
  S.~W.,  {Forman} W.,  {Jones} C.,   {Kaastra} J.~S.,  2009, \mn@doi [\mnras]
  {10.1111/j.1365-2966.2009.14860.x}, \href
  {http://adsabs.harvard.edu/abs/2009MNRAS.398...23W} {398, 23}

\bibitem[\protect\citeauthoryear{{Werner} et~al.,}{{Werner}
  et~al.}{2013}]{Werner:2013}
{Werner} N.,  et~al., 2013, \mn@doi [\apj] {10.1088/0004-637X/767/2/153}, \href
  {http://adsabs.harvard.edu/abs/2013ApJ...767..153W} {767, 153}

\bibitem[\protect\citeauthoryear{{Werner} et~al.,}{{Werner}
  et~al.}{2014}]{Werner:2014}
{Werner} N.,  et~al., 2014, \mn@doi [\mnras] {10.1093/mnras/stu006}, \href
  {http://adsabs.harvard.edu/abs/2014MNRAS.439.2291W} {439, 2291}

\bibitem[\protect\citeauthoryear{{Yang} \& {Reynolds}}{{Yang} \&
  {Reynolds}}{2016}]{Yang:2016}
{Yang} H.-Y.~K.,  {Reynolds} C.~S.,  2016, \mn@doi [\apj]
  {10.3847/0004-637X/829/2/90}, \href
  {http://adsabs.harvard.edu/abs/2016ApJ...829...90Y} {829, 90}

\bibitem[\protect\citeauthoryear{{Zhuravleva}, {Allen}, {Mantz}  \&
  {Werner}}{{Zhuravleva} et~al.}{2017}]{Zhuravleva:2017}
{Zhuravleva} I.,  {Allen} S.~W.,  {Mantz} A.~B.,   {Werner} N.,  2017,
  preprint, \href {http://adsabs.harvard.edu/abs/2017arXiv170702304Z} {}
  (\mn@eprint {arXiv} {1707.02304})

\makeatother
\end{thebibliography}


\label{lastpage}
\end{document}